\documentstyle[11pt,epsfig]{article}
\title{
 Polarized Structure of Nucleon In the Valon Representation
}

\author{Firooz Arash$^{(a)}$\footnote{e-mail: farash@cic.aut.ac.ir} and
Fatemeh Taghavi-Shahri$^{(b)}$\\
$^{(a)}$ Physics Department, Tafresh University, Tafresh, Iran \\
$^{(b)}$
Physics Department, Iran University of Science and Technology, Narmak, Iran \\
}
\date{\today}
\begin{document}
\maketitle
\begin{abstract}
We have utilized the concept of {\it{valon}} model to calculate
the spin structure functions of proton, neutron, and deuteron. The
valon structure itself is universal and arises from the
perturbative dressing of the valence quark in QCD. Our results
agree rather well with all of the relevant experimental data on
$g_{1}^{p,n,d}$ and $\frac{g_{A}}{g_{V}}$, and suggests that the
sea quark contribution to the spin of proton is consistent with
zero. It also reveals that while the total quark contribution to
the spin of a valon, $\Delta \Sigma_{valon}$, is almost constant
at $Q^{2}\geq 1$ the gluon contribution grows with the increase of
$Q^2$ and hence requiring a sizable negative orbital angular
momentum component $L_{z}$. This component along with the singlet
and non-singlet parts are calculated in the Next-to-Leading order
in QCD . We speculate that gluon contribution to the spin content
of the proton is about $60\%$ for all $Q^2$ values. Finally, we
show that the size of gluon polarization and hence, $L_{z}$, is
sensitive to the initial scale $Q_{0}^{2}$.

\end{abstract}
\section{INTRODUCTION}
A central goal in the study of QCD is to understand the structure
of hadrons in terms of their quark and gluon degrees of freedom.
The most direct tool and sensitive test for probing the quark and
gluon substructure of hadrons is the polarized Deep Inelastic
Scattering (DIS)processes. In such experiments detailed
information can be extracted on the shape and magnitude of the
spin dependent parton distributions, $\delta q_{f}(x,Q^{2})$. Deep
inelastic scattering reveals that the nucleon is a rather
complicated object consisting of an infinite number of quarks,
anti-quarks, and gluon. It is a common belief that other strongly
interacting particles also exhibit similar internal structure.
However, under certain conditions, hadrons behave as if they were
composed of three (or two) constituents. Examples are the magnetic
moments of the baryons, meson and baryon spectroscopy, the
meson-baryon couplings and the ratio of total cross sections such
as $\frac{\sigma(\pi N)}{\sigma(NN)}$ and so on. Thus, it seems to
make sense to decompose a nucleon into three constituent quarks
called U and D. They would carry the internal quantum numbers of
the nucleon. On the other hand, in DIS one observes that a nucleon
has a composition of essentially an infinite number of
quark-antiquark pairs and gluons, in addition to its valence
quarks. One might identify the valence quark with a constituent
quark, but this would imply that the three quark picture is a very
rough approximation and both $q\bar{q}$ pair and gluon degrees of
freedom need to be added to the picture. In doing so, it would be
very difficult to understand why the three quark picture of a
baryon works so well in many circumstances. One way of reconciling
this apparent contradiction is to consider a constituent quark as
quasi-particle with a non-trivial internal structure of its own;
consisting of a valence quark and a sea of $q\bar{q}$ pairs and
gluons. Such an interpretation of constituent quark is not new,
more that 30 years ago it was advocated by Altarelli and Cabibo
\cite{1}. R.C. Hwa developed a more elaborated version by
introducing the so called {\it{valon model}} \cite{2} (the term
which we will use hereafter) and applied it to a variety of
phenomena with great success. More recent indication for the
existence of the valon can be inferred from the measurements of
the Natchmann moments of the proton structure functions at
Jefferson Laboratory. They point to the existence of a new scaling
that can be interpreted as a constituent form factor consistent
with the elastic nucleon data \cite{3}. This finding suggests that
the proton structure originates from elastic coupling with
extended objects inside the proton. In References
\cite{4}\cite{5}\cite{6}, the valon concept is utilized to
calculate the unpolarized structure function of a number of
hadrons. The results are in excellent agreement with the
experimental data. Altarelli has also calculated the pion
structure function in the constituent quark representation
\cite{7}, using a deconvolution procedure. On a more theoretical
front, M. Lavelle and D. McMullan \cite{8} \cite{9} proved that
one can dress a QCD Lagrangian field to all orders in perturbation
theory and construct a constituent quark in conformity with the
color confinement. From this point of view, a valon is defined as
a structureful object emerging from the dressing of a valence
quark with gluons and $q\bar{q}$ pairs in QCD. Chiral Models in
the realm of non-perturbative QCD also require the dressing of a
valence quark and thus producing a structureful object. These
results and the success of the valon model in describing the
unpolarized structure of hadrons and a number of other low $P_T$
hadronic phenomena lend credit for the study of hadrons in the
valon representation. \newline In this paper we calculate the
polarized structure of a valon and extract the polarized structure
function, $g_1$, of the nucleon and the deuteron and compare the
results with the experimental data. A number of similar attempts
are also made to derive the nucleon spin from the quark models
\cite{10},\cite{11}, \cite{12}, \cite{13}, \cite{14}. Our model
differs from those in that we calculate the polarized structure of
a constituent quark (the valon) directly from QCD processes in the
Next-to-leading and investigate its peculiarities and distinctive
features. In order to be clear, we define a valon as a valence
quark plus its associated sea partons, emerging from dressing
processes. In a bound state problem those processes are virtual
and a good approximation for the problem is to consider a valon as
one integral unit whose internal structure cannot be resolved.
Therefore, it is assumed that the spin of the nucleon is provided
by the combination of the spins of the valons. In a scattering
situation, on the other hand, the virtual partons inside a valon
can be excited and be put on mass shell. It is therefore more
appropriate to think of a valon as a cluster of partons with some
momentum and helicity distribution. If the valon has a non-trivial
internal structure then the question arises whether its spin
structure is also complex as it seems to be the case for the
nucleon. This issue will also be addressed. In Reference \cite{15}
this model is used to calculate polarized structure functions.
While obtaining results that are in agreement with the
experimental data, however, it contains misleading and at points
even counter intuitive ingredients. We will address them
throughout this paper.
\newline The organization of the paper is as follows: First we
will outline the formalism for calculating the spin structure of
the valon, then the polarized structure function of the nucleon
and deuteron will be evaluated and it will be shown that the
orbital angular momentum of partons in a valon plays a central
role in describing the spin of nucleon.

\section{Polarized Valon Structure }
In this section we will utilize the extended work done on the
development of NLO calculation of the moments, to evaluate the
polarized structure of a valon. We should stress that this is not
a new next-to-leading order calculation, but it is an exploration
of the existing calculations in the valon framework.\\
By definition, a valon is a universal building block for every
hadron; that is, its structure is independent of the hosting
hadron. The valons play a dual role in hadrons: (i) they interact
with each other in a way that is characterized by the valon wave
function and (ii) they respond independently in an inclusive hard
collision with a $Q^{2}$ dependence that can be calculated in QCD
at high $Q^{2}$. In role (i) they are the constituents of bound
state problem involving the confinement at large distances. In
role (ii) they are quasi-particles whose internal structure are
probed with high resolution and are related to the short distance
problem of current operators. This picture suggests that the
structure function of a hadron involves a convolution of two
distributions: valon distribution in the hadron and the parton
distribution in the valon. In an unpolarized situation we may
write:
\begin{equation}
F^{h}_{2}(x,Q^{2})=\sum_{\it{valon}}\int_{x}^{1} dy
G_{\it{valon}}^{h}(y) F^{\it{valon}}_{2}(\frac{x}{y},Q^{2})
\end{equation}
where $F^{\it{valon}}_{2}(\frac{x}{y},Q^{2})$ is the structure
function of the probed valon and can be calculated in Perturbative
QCD to a certain degree of approximation. If $Q^{2}$ is small
enough we may identify $F^{\it{valon}}(x, Q^{2})$ as $\delta(z-1)$
at some point, for the reason that we cannot resolve its internal
structure at that $Q^{2}$ value. Similarly, for a polarized hadron
we can write
\begin{equation}
g^{h}_{1}(x,Q^{2})=\sum_{\it{valon}}\int_{x}^{1}\frac{dy}{y}
\delta G_{\it{valon}}^{h}(y) g^{\it{valon}}_{1}(\frac{x}{y},Q^{2})
\end{equation}
where $\delta G_{\it{valon}}^{h}(y)$ is the helicity distribution
of the valon in the hosting hadron and
$g^{\it{valon}}_{1}(\frac{x}{y}, Q^{2})$ is the polarized
structure function of the valon. At high $Q^{2}$ for a U-type
valon one can write $g^{\it{valon}}$ as follows:
\begin{equation}
2 g_{1}^{U}(z,Q^{2})=\frac{4}{9}(\delta G_{\frac{u}{U}}+ \delta
G_{\frac{\bar{u}}{U}}) +\frac{1}{9}(\delta G_{\frac{d}{U}}+\delta
G_{\frac{\bar{d}}{U}}+\delta G_{\frac{s}{U}}+\delta
G_{\frac{\bar{s}}{U}}) +...
\end{equation}
where all the functions on the right-hand side are the helicity
functions for finding polarized quarks with momentum fraction $z$
in a U-type valon at that $Q^{2}$. These functions or certain
combinations of them can be calculated in QCD and assumed to be
known as we will meet them later. Similar expression can also be
written for the D-type valon. To describe the polarized parton
distribution inside a valon , we will work in the moment space,
where the moment of the polarized parton distribution in a valon
is defined as:
\begin{equation}
\Delta f (n,Q^2)=\int^{1}_{0} z^{n-1} \delta f(z,Q^2)dz.
\end{equation}
$\delta f(z,Q^2)$ corresponds to parton helicity densities in a
valon. The moments of the valon structure function are expressed
completely in terms of $Q^{2}$ through the evolution parameter
$t$:
\begin{equation}
t={\it{ln}}\frac{\it{ln}\frac{Q^2}{\Lambda^2}}{\it{ln}\frac{Q_{0}^{2}}{\Lambda^2}}.
\end{equation}
We work in ${\overline{MS}}$ scheme where
$\Lambda_{QCD}^{\overline{MS}}$ is given by
\begin{equation}
\Lambda_{QCD}^{\overline{MS}}=\mu
Exp\{-\frac{1}{2}[\frac{1}{\beta_{0}
\alpha_{s}(\mu^2)}-\frac{\beta_1}{\beta_{0}^{2}}log(\frac{1}{\beta_{0}
\alpha_{s}(\mu^2)}+\frac{\beta_{1}}{\beta_{0}})]\},
\end{equation}
where $\mu$ is the factorization scale and the $\beta$ functions
are as follows
\begin{equation}
\beta_{0} = \frac{11}{3}C_{A} -f\frac{4}{3}T_{F}, \hspace{2cm}
\beta_{1}=\frac{34}{3}C_{A}^{2} -f\frac{20}{3}C_{A} T_{F}-4f
C_{F}T_{F}.
\end{equation}
Here $C_{A}=3$, $T_{F}=\frac{f}{2}$, $C_{F}=\frac{4}{3}$, and $f$
is the number of active flavors. A NLO fit to the $g_{1}/F_{1}$
with massless quarks is performed in \cite{16} and favors a value
$\Lambda_{QCD}=0.235 \pm 0.035$ $GeV$. This is very close to our
choice of $\Lambda_{QCD}=0.22$ $GeV$ for the unpolarized case. we
will maintain this value along with
$Q_{0}^2=0.283$ $GeV^2$ as in \cite{4}.\\
 Moments of the polarized valence and sea quarks in a polarized valon
are:
\begin{equation}
\delta M_{\frac{\delta q_{v}}{valon}} = \delta M_{NS}(n, Q^{2})
\end{equation}
\begin{equation}
\delta M_{\frac{\delta q_{sea}}{valon}}=\frac{1}{2f}(\delta
M_{S}-\delta M_{NS})(n,Q^2)
\end{equation}
where $f$ is the number of flavors and $\delta M_{S,NS}$ are
polarized singlet and non-singlet moments defined as:
\begin{equation}
\delta M_{NS
\pm}(n,Q^{2})=\{1+\frac{\alpha_{s}(Q^2)-\alpha_{s}(Q_{0}^{2}}{2
\pi}(\frac{-2}{\beta_{0}})(\delta {\bf{P}}^{(1)n}_{NS
\pm}-\frac{\beta_{1}}{2 \beta_{0}}\delta
{\bf{P}}^{(0)n}_{qq})\}{\bf{L}}^{-(\frac{2}{\beta_{0}})\delta
P^{(0)n}_{qq}}
\end{equation}
\begin{eqnarray}
 \left( \begin{array}{c}
 \delta M_{S}(n,Q^{2}) \\ \delta M_{G}(n,Q^{2})
 \end{array} \right)
=\{{\bf{L}}^{-(\frac{2}{\beta_{0}})\delta \hat{P}^{(0)n}}+
\frac{\alpha_{s}(Q^2)}{2 \pi}{\bf{\hat{U}
L}}^{-(\frac{2}{\beta_{0}})\delta
\hat{P}^{(0)n}}-\frac{\alpha_{s}(Q_{0}^{2})}{2
\pi}L^{-(\frac{2}{\beta_{0}})\delta
\hat{P}^{(0)n}}{\bf{\hat{U}}}\}\left( \begin{array}{c}
 {1} \\{0}
 \end{array}\right)
\end{eqnarray}
The column matrix on the right hand side describes our initial
input densities and they constitute an essential part of this
work. We have inferred them from the properties of the model: The
valon structure function has the property that it becomes
$\delta(z-1)$ as $Q^2$ is extrapolated to $Q^{2}_{0}$ (beyond the
region of validity). This mathematical boundary condition means
that the internal structure of the valon cannot be resolved at
$Q^{2}_{0}$ in the NLO approximation. Consequently, when this
property is applied to Eq. (2), the structure function of the
nucleon becomes directly related to $x\delta G_{valon}^{h}(x)$ at
that value of $Q^{2}_{0}$; that is, $Q^{2}_{0}$ is the
leading-order effective value at which the hadron can be regarded
as consisting of only three (two) valons for baryons(mesons). In
the moment space it is the Mellin transform of the
$\delta$-function, being equal to one, that enters. Naturally,
then $\delta M_{g}(n,Q^{2}_{0})=0$ and it is reasonable to set
$\delta f(z,Q_{0}^{2})= f(z,Q_{0}^{2})$, for the quark sector. As
$Q^{2}$ increases beyond a small enough value, say $Q^{2}_{v}$,
where we may identify valon structure as $\delta(z-1)$, one
expects the valon structure to develop a tail in the $0<z<1$
region due to gluon radiation. Reliable calculations are only
possible for higher $Q^2$ values; let it be for $Q^2 > Q_{1}^{2}$.
Between $Q^{2}_{v}$ and $Q^{2}_{1}$ higher twist terms are
involved and the picture is more complicated, but that is also the
region where the most important part of $Q^{2}$ evolution takes
place. In reference [15] the initial moments of both gluon and
singlet sectors are set equal to one. Such a choice has no
justification in the valon model. Moreover, it is known that a
fully saturated initial input density for the gluon is disfavored
[16]. Basically, initial input densities are determined from the
experimental data, but for the valon there is no experimental
data, therefore our choice of initial input densities, based on
the mathematical conditions of the model does not invalidate it,
nor does it violate the positivity constraint. \newline In
Eqs.(10,11), ${\bf{L}} \equiv
\alpha_{s}(Q^2)/\alpha_{s}(Q^{2}_{0})$, and $\delta
\hat{P}^{(0)n}$ is $2\times 2$ singlet matrix of splitting
functions, given by
\begin{eqnarray}
\delta \hat{P}^{(0)n}= \left ( \begin{array} {c} \delta
P^{(0)n}_{qq} \hspace{0.75cm} 2f\delta P^{(0)n}_{qg} \\ \delta
P^{(0)n}_{gq}\hspace{0.75cm}\delta P^{(0)n}_{gg}
\end{array} \right ),
\end{eqnarray}
where $\delta P^{(0)n}_{lm}$ are the $n^{th}$ moments of the
polarized splitting functions and ${\bf{U}}$ accounts for the
2-loop contributions as an extension to the leading order. The
explicit forms of these functions are given in \cite{17} in the
next-to-leading order. Now it is straightforward to calculate the
moments of polarized partons inside a valon at any $Q^{2}$ or $t$
value. These moments are shown in Figure 1.
\begin{figure}
\centerline{\begin{tabular}{ccc}
\epsfig{figure=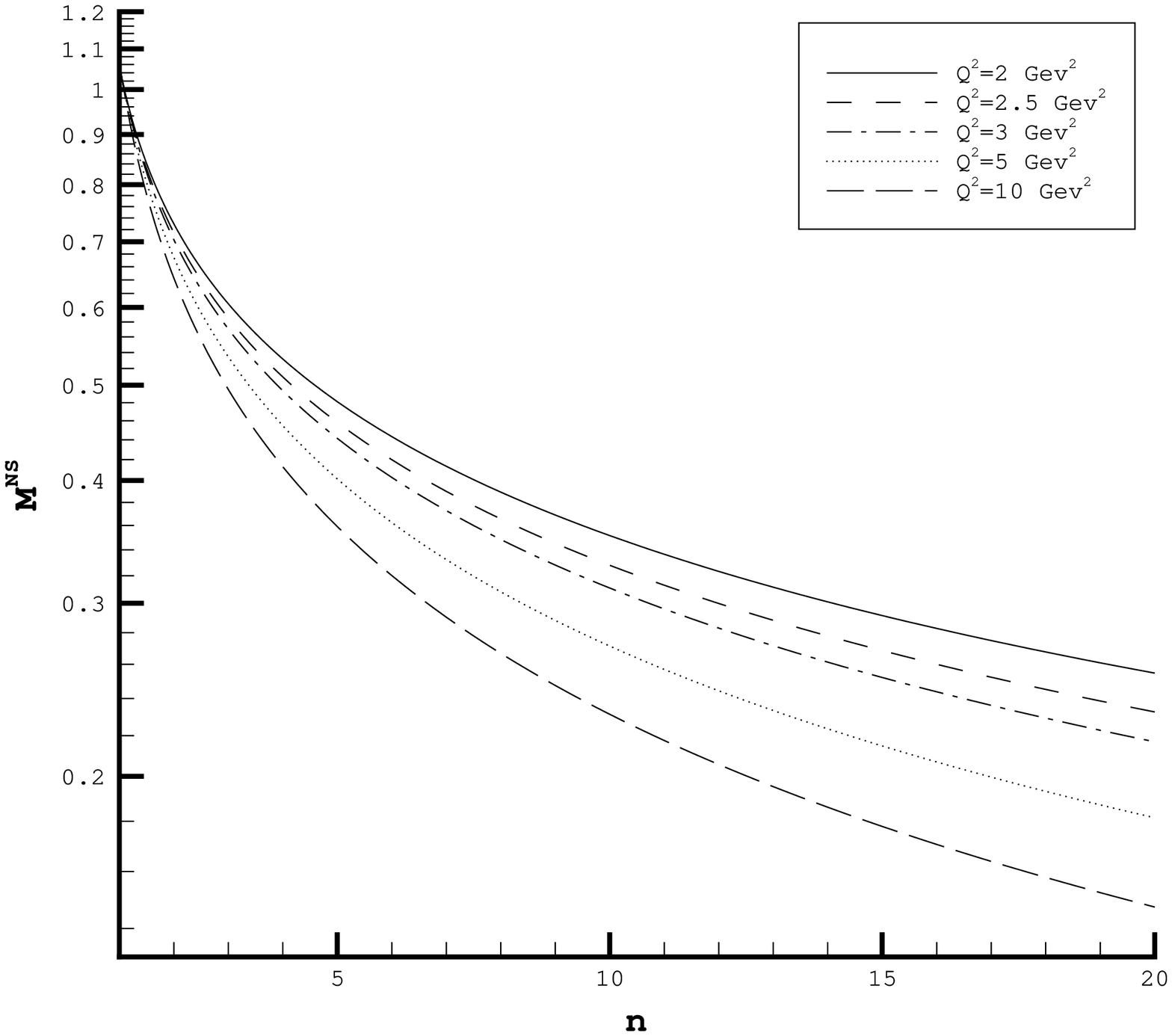,width=6cm}
\epsfig{figure=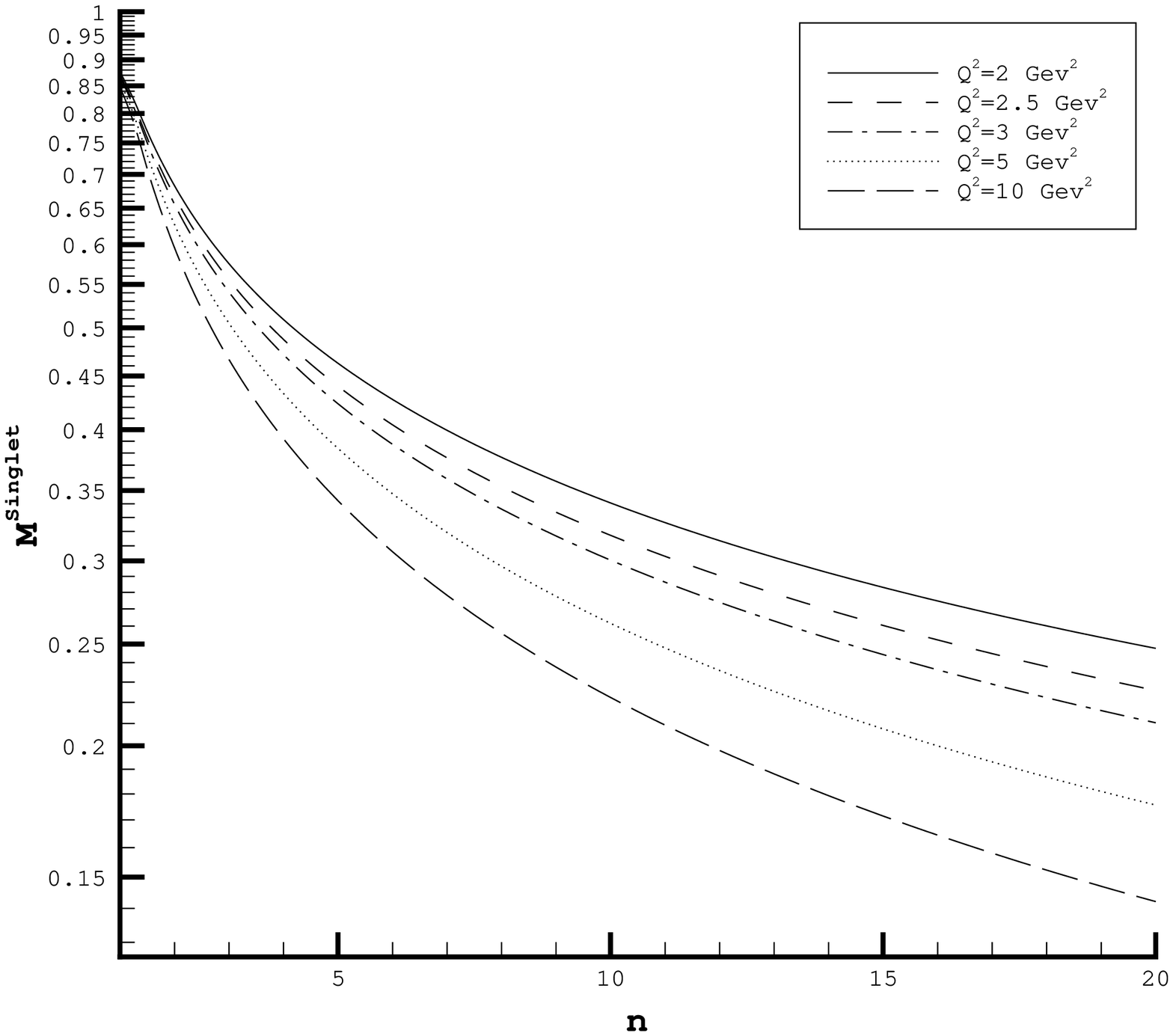,width=6cm}
\epsfig{figure=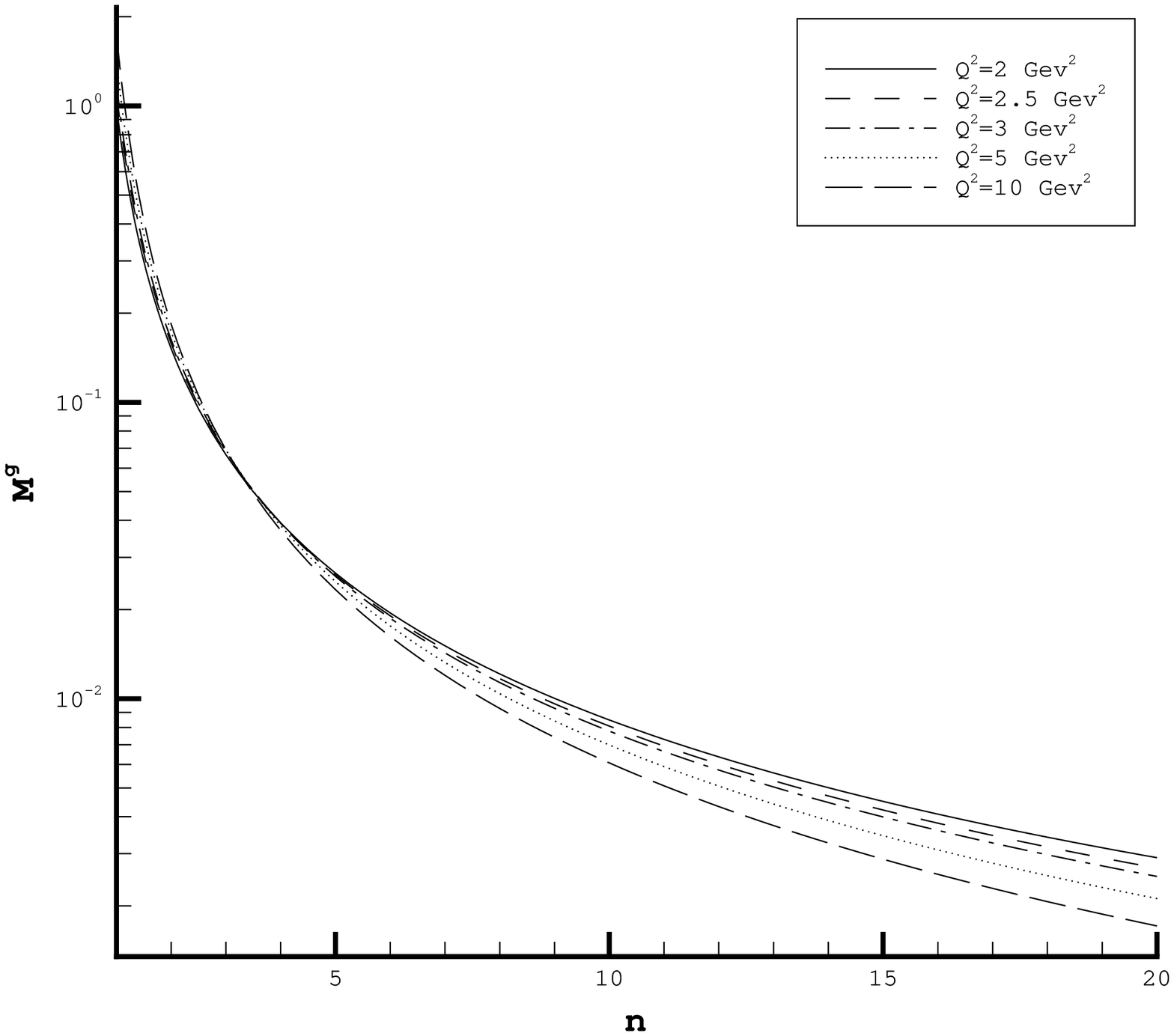,width=6cm}
\end{tabular}}
\caption{\footnotesize Moments of singlet, non-siglet quark, and
gluon distributions in a valon at several $Q^2$ values.}
\label{fig1}
\end{figure}
The $z$-dependence of the polarized parton distributions is
obtained by utilizing the usual inverse Mellin transformation.
\begin{equation}
\delta q_{NS,S,G}^{valon}(z,Q^{2})=\frac{1}{\pi}\int^\infty_0
{\it{Im}}[e^{i\phi} z^{-c-we^{i\phi}}\delta
M^{NS,S,G}(n=c+we^{i\phi},Q^2)] dw,
\end{equation}
where NS, S, and G stand for non-singlet, singlet and gluon,
respectively. In what follows we shall only be interested in
quantities averaged over $z$ and thus in the $n=1$ moments. The
first moments are defined by
\begin{equation}
\Delta f(Q^2)=\int dz \delta f(z,Q^2)
\end{equation}
\begin{figure}
 \epsfig{figure=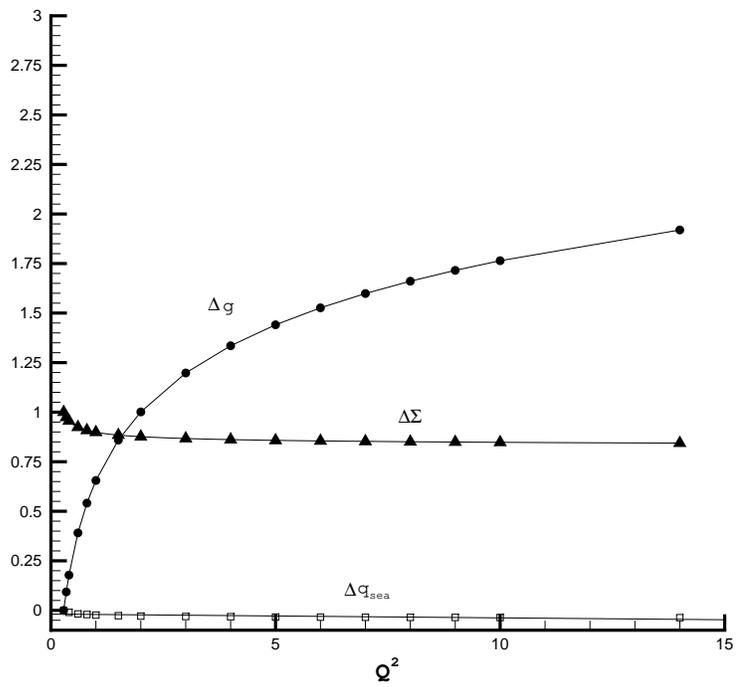,width=12cm}
\caption{\footnotesize First Moments, $\Delta g(n=1, Q^2)$,
$\Delta \Sigma(n=1, Q^2)$, and
 $\Delta q_{sea}(n=1, Q^2)$ of various components in a valon as a function of $Q^2$ }
\label{figure 2.}
\end{figure}
There is a simple physical interpretation for the quantities like
$\Delta \Sigma$, $\Delta q$ and $\Delta g$ : they are related to
the total $z$ component of quark and gluon spins, thus
\begin{equation}
<S_{z}>_{q}=\frac{1}{2}\Delta q,\hspace{3cm} <S_{z}>_{g}=\Delta g
\end{equation}
These quantities for the valon are shown in Figure 2. The results
imply that the total quark contribution to the spin of a valon,
decreases from 1 to $\Delta \Sigma=0.88$, in the range of
$Q^{2}=[0.283,1]$ $GeV^{2}$ and remains almost independent of
$Q^{2}$ thereafter. In other words, if the valon consisted only of
quarks (valence plus sea) it would have been enough to account for
$\simeq 90\%$ of the valon spin at $Q^{2}\geq 1$. Evidently,
however, there is a sizable gluon component, $\Delta g$, which
increases with $Q^{2}$ as shown in Figure 2. In figure 3 we have
shown the variation of $\Delta q_{sea}$ and $\Delta q_{valence}$
as a function of $Q^{2}$. The variation of $\Delta q_{sea}$ and
$\Delta \Sigma$ with $Q^{2}$ is very marginal. we have checked
that $\Delta q_{valence}=[1, 1.08]$ for the range of $Q^2=
[0.283,10^6]$ $GeV^2$; whereas $\Delta q_{sea}$ varies from 0 to
-0.043 for the same range of $Q^2$. This weakly $Q^{2}$ dependent
behavior of $\Delta q_{valence}$ and $\Delta q_{sea}$, are well
understood: in the leading order one expects the total quark spin
to be constant due to the vanishing of quark anomalous dimensions
at $n=1$. In the Next-to-Leading Order, however, they are
marginally $Q^{2}$ dependent due to $\Delta P^{(1)}_{NS}\neq 0$.
The fact that sea quark polarization in the valon is consistent
with zero can  also be understood on theoretical grounds. The
valon structure is generated by perturbative dressing in QCD. In
such processes with massless quarks, helicity is conserved and
therefore, the hard gluons cannot induce sea polarization
perturbatively. It is also worth to note that $\Delta
q_{sea}\simeq 0$ is in good agreement with HERMES data \cite{18} \cite{19}. \\
\begin{figure}
\centerline{\begin{tabular}{ccc}
\epsfig{figure=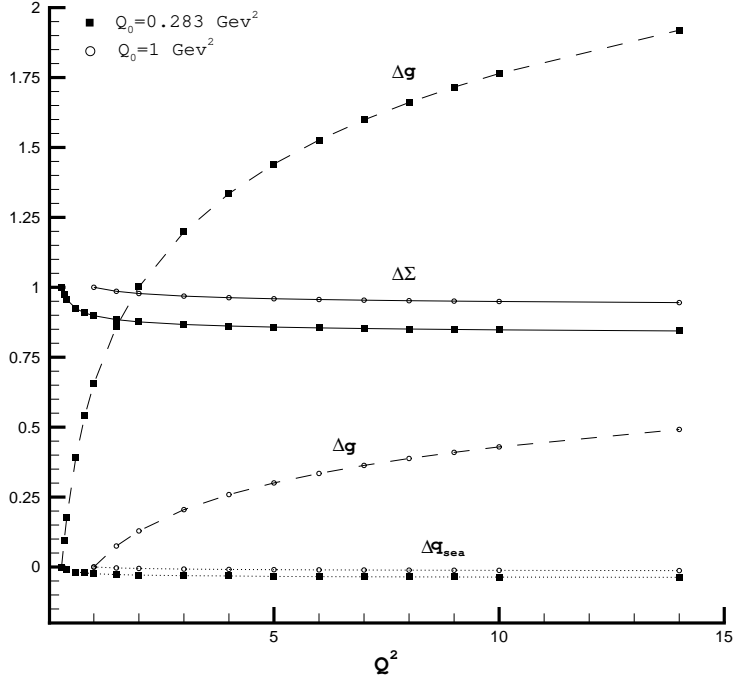,width=12cm}
\end{tabular}}
\caption{\footnotesize variations of $\Delta q_{valence}$ and
$\Delta q_{sea}$ in a valon as a function of $Q^2$. Full squares
are the results calculated with $Q_{0}^{2}=0.283$ $GeV^{2}$ and
open circles correspond to $Q_{0}^{2}=1$ $GeV^{2}$.} \label{fig3}
\end{figure}
Had we chosen $Q_{0}^{2}=1$ $GeV^{2}$, we would have obtained
$\Delta \Sigma=1$ for all values of $Q^{2}$ with little change in
$\Delta q_{sea}$, and with a much reduced gluon polarization, as
compared to our results with $Q_{0}^{2}=0.238$ $GeV^{2}$ as can be
seen in Figure 3. Such a choice with initial gluon helicity
distribution equal to zero, however, is inconsistent with the
mathematical condition of the model. The use of a different
initial gluon helicity distribution, instead of zero, also would
have been a pure guess and at best would require data fittings
that we have avoided. \\
It is obvious that due to large gluon polarization in a valon,
these results do not add up to give the spin $\frac{1}{2}$ of the
valon and do not satisfy the sum rule
$\frac{1}{2}=\frac{1}{2}\Delta\Sigma + \Delta g$. The gluon
contribution to the valon spin grows as $Q^2$ increases, while
$\Delta \Sigma$ remains almost unchanged beyond $Q^{2}=1$
$GeV^{2}$. It is the present wisdom that the above sum rule must
be replaced with a more realistic one
\begin{equation}
\frac{1}{2}=\frac{1}{2}\Delta \Sigma + \Delta g +L_{z}
\end{equation}
where, $L_{z}$ is the orbital angular momentum carried by the sea
partons ($q-\bar{q}$ pairs and gluons) within the valon. The size
of this orbital angular momentum turns out to be large and
negative, mainly competing with the gluon contribution. Ratcliffe
\cite{20} was the first to point out the necessity of including
orbital angular momentum dependence of the evolution equation and
predicted a negative value for $<L>_{z}$ of the sea partons in the
proton. In Fig. 4 we present the Next-to-Leading Order result for
the orbital angular momentum in a valon, $\L_{z}^{valon}(Q^2)$. A
leading order calculation is given in \cite{21}.
\begin{figure}
\epsfig{figure=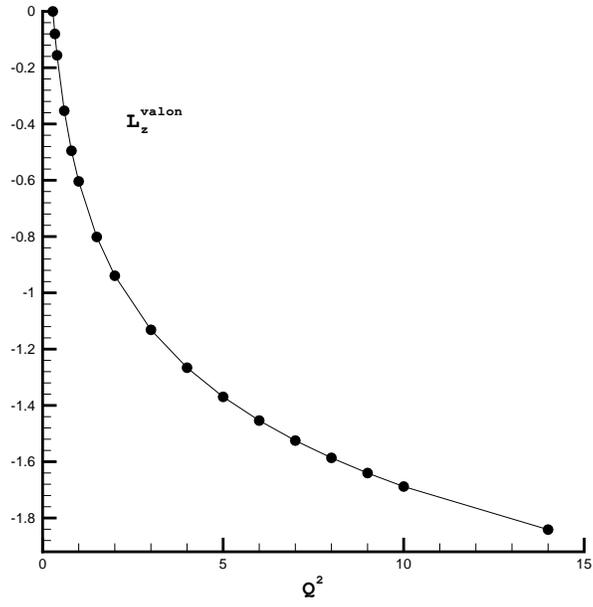,width=12cm} \caption{\footnotesize Orbital
angular momentum, $L_{z}^{valon} (Q^2)$, component of partons in a
valon as a function of $Q^2$. The results corresponds to
$Q_{0}^{2}=0.283$ $GeV^{2}$.} \label{figure 4.}
\end{figure}
The existence of non-zero internal orbital momentum in the valon
implies that there are substantial correlations among partons. It
has been argued that the presence of quark pairs inside hadron
resembles superconductivity \cite{22}. An extension of the theory
of superconductivity to the anisotropic case \cite{23}, shows that
the presence of anisotropy leads to axial symmetry of pairing
correlations around the anisotropy direction and to the particle
currents induced by the pairing correlations. The particle number
conservation, then requires that the cloud of correlated particles
to rotate around the central particle in opposite direction, the
so-called {\it{Backflow}} current. This pairing induced orbital
angular momentum is proportional to the density of correlated
particles. The analogy seems to match the valon picture, where the
internal structure of the valon originates entirely from QCD
processes.
\section{Polarized Nucleon Structure Function}
In the previous section the polarized structure of a valon is
completely specified. We are now in a position to carry forward
and investigate the implications of the model at the hadronic
level. Our starting point is Eq.(2), where the only unknown
element is the valon helicity distribution, $\delta
G_{\it{valon}}^{h}(y)$, since the valon structure function
$g_{1}^{valon}$ is now completely given. In the analysis of
\cite{21}, for the leading order, we assumed that the polarized
valon distribution is related to the, by now well determined,
unpolarized valon distribution via:
\begin{equation}
\delta G_{j}(y) = \delta F_{j}(y) G_{j}(y)
\end{equation}
where, $G_{j}(y)$ is the unpolarized valon distribution of $j=U,
D$ kinds. $G_{j}(y)$ are given in \cite{4} \cite{5} \cite{6} for a
variety of hadrons. They mimic the hadron wave functions.
\newline In the absence of experimental knowledge on $\delta
G_{\it{valon}}^{h}(y)$, the safe way to determine them is to fit
the experimental values of $g_{1}^{p}$ at some $Q^2$ with the form
given by
\begin{equation} \delta F_{j}(y, Q^{2}_{0})=
 N_{j}y^{\alpha_{j}}(1-y)^{\beta_{j}}(1+ a_{j} y^{0.5} + b_j y +c_j y^{1.5} +d_j y^2)
\end{equation}
where, $j$ stands for $U$ and $D$ type valon and parameters
$\alpha_{j}$, $\beta_{j}$, etc. are given in Table I.
Consequently, the polarized valon distributions in a proton are
now completely specified and are given by:
\begin{equation}
\delta G_{\frac{U}{P}}(y)=\delta F_{U} G_{\frac{U}{P}}\hspace{2cm}
\delta G_{\frac{D}{P}}(y)=\delta F_{D} G_{\frac{D}{P}}
\end{equation}
\\
\\
{\it{Table I. Numerical values of the parameters in Eq. (18).}} \\
\begin{tabular}{ccccccccc}
\hline\hline $valon (j)$ & $N_{j}$ & $\alpha_{j}$ &
$\beta{_j} $ & $a_j $ & $b_j $ & $c_j $ & $d_j$ \\
\hline
$U$ & 3.44 & 0.33 & 3.58 & -2.47 & 5.07 & -1.859 & 2.780 \\
$D$ & -0.568 & -0.374 & 4.142 & -2.844 & 11.695 & -10.096 & 14.47
\\
\hline\hline
\end{tabular}
\\ \\
Now we can calculate the polarized hadronic structure function,
$g_{1}^{h}$. For this purpose all that we need is to substitute
$\delta G_{\frac{U}{P}}(y)$ and $\delta G_{\frac{D}{P}}(y)$ from
Eq. (19) into Eq. (2) and perform the convolution integral. In
Fig. 5 we present the results for proton, $g_{1}^{p}$, and compare
them with the experimental data \cite{18},\cite{19}
\cite{24},\cite{25}, and with the calculations of \cite{16},
\cite{26}, and \cite{27}.

\begin{figure}
\centerline{\begin{tabular}{cccccccc}
\epsfig{figure=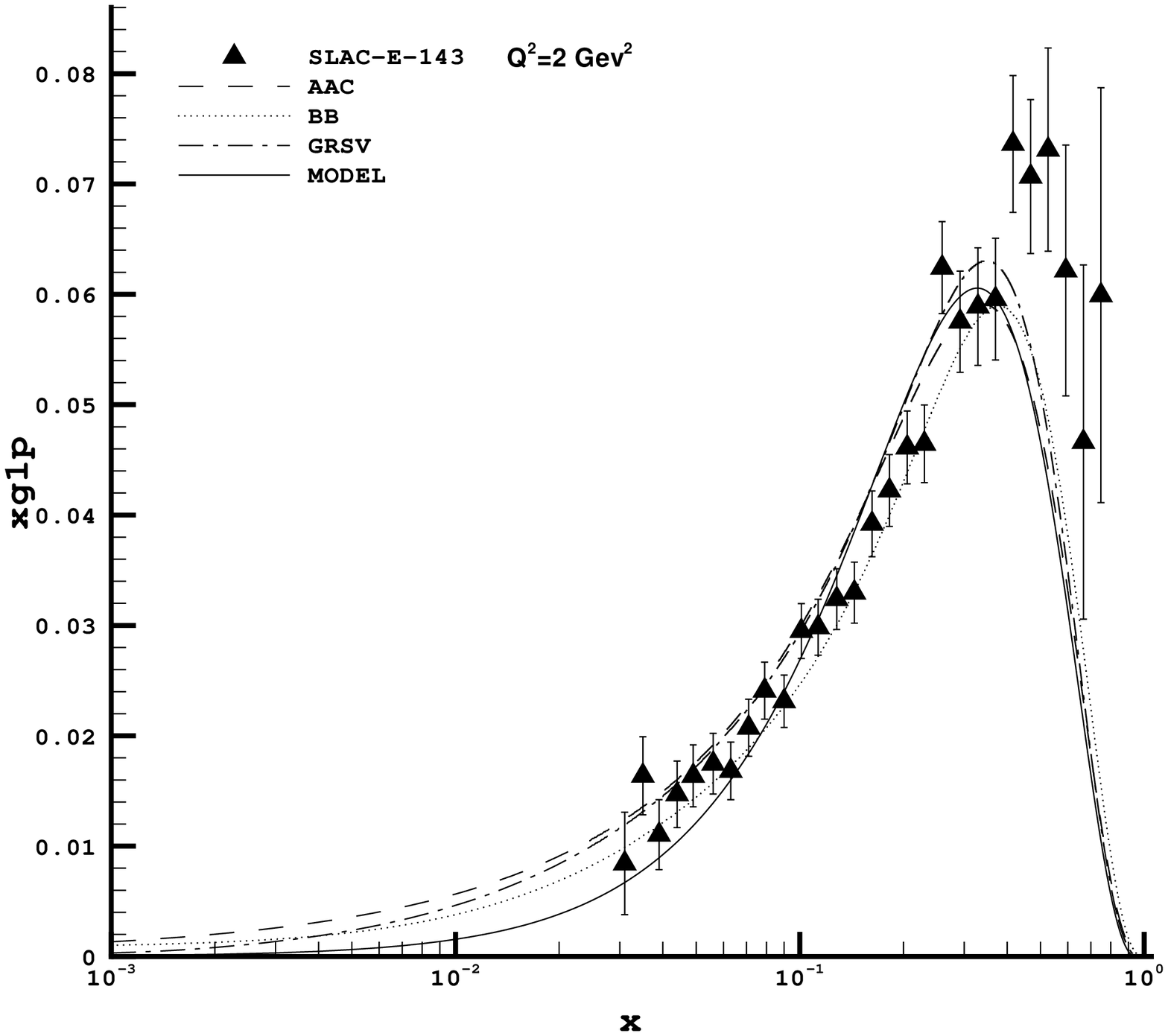,width=9cm}
\epsfig{figure=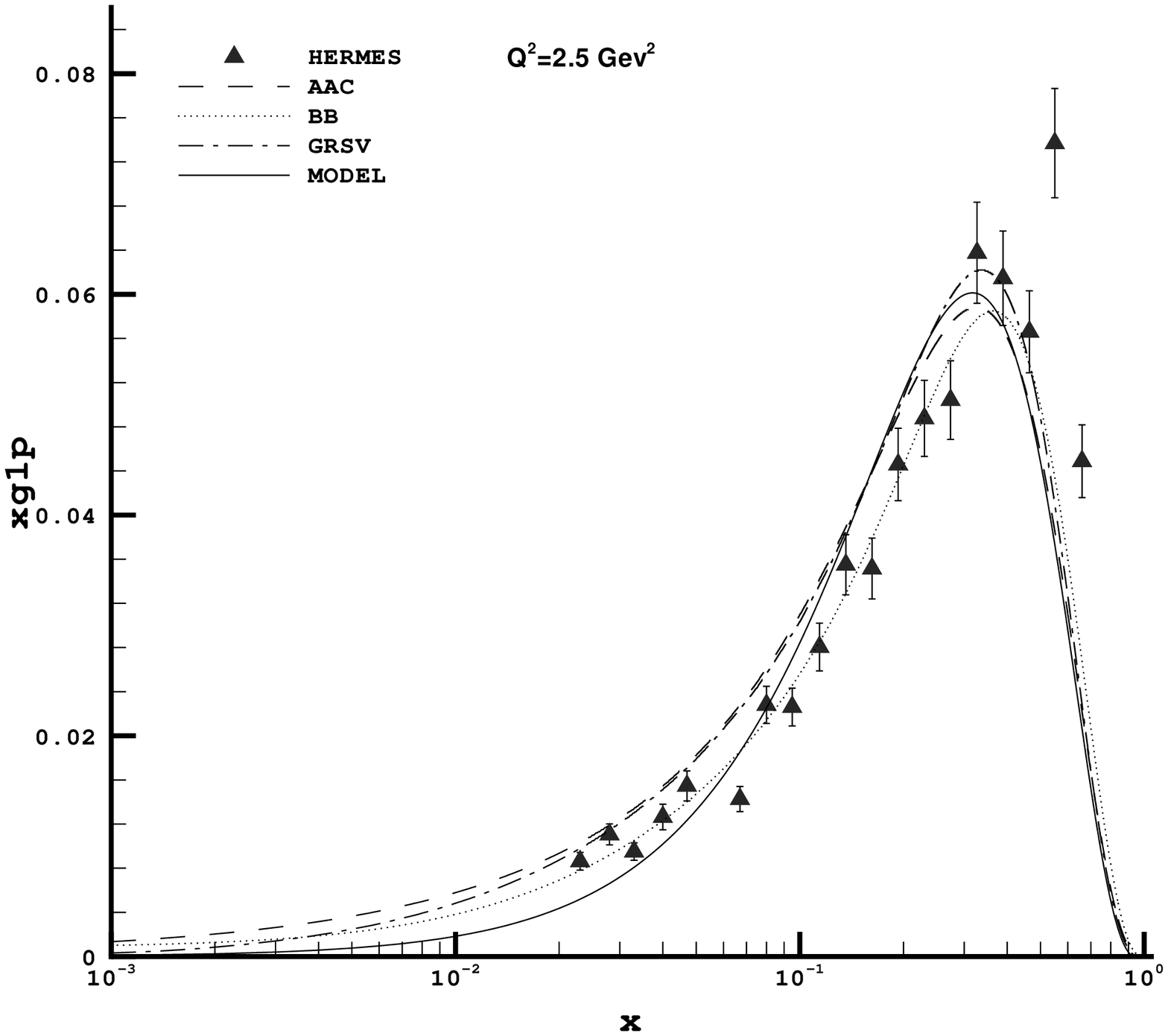,width=9cm} \\
\epsfig{figure=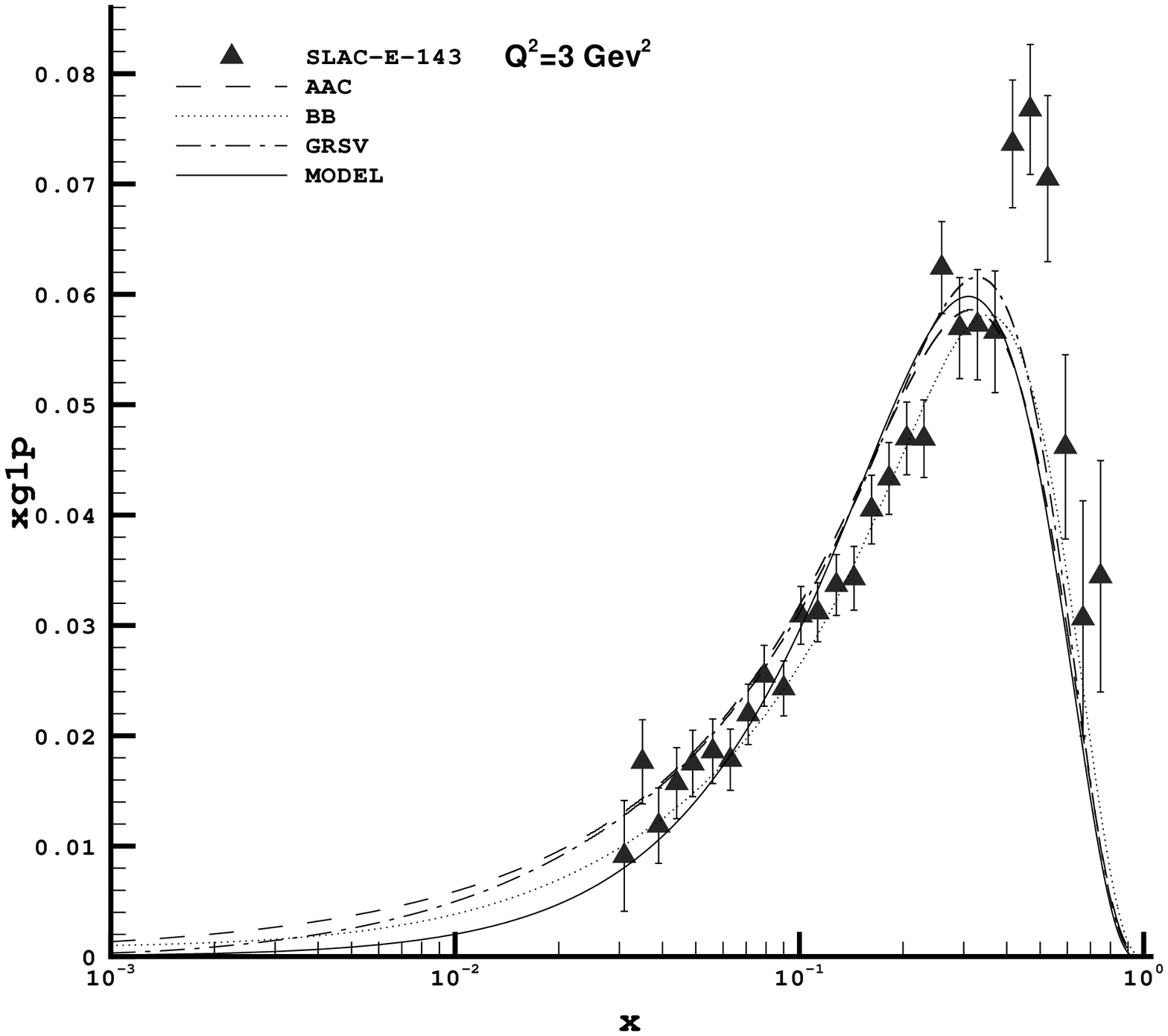,width=9cm}
\epsfig{figure=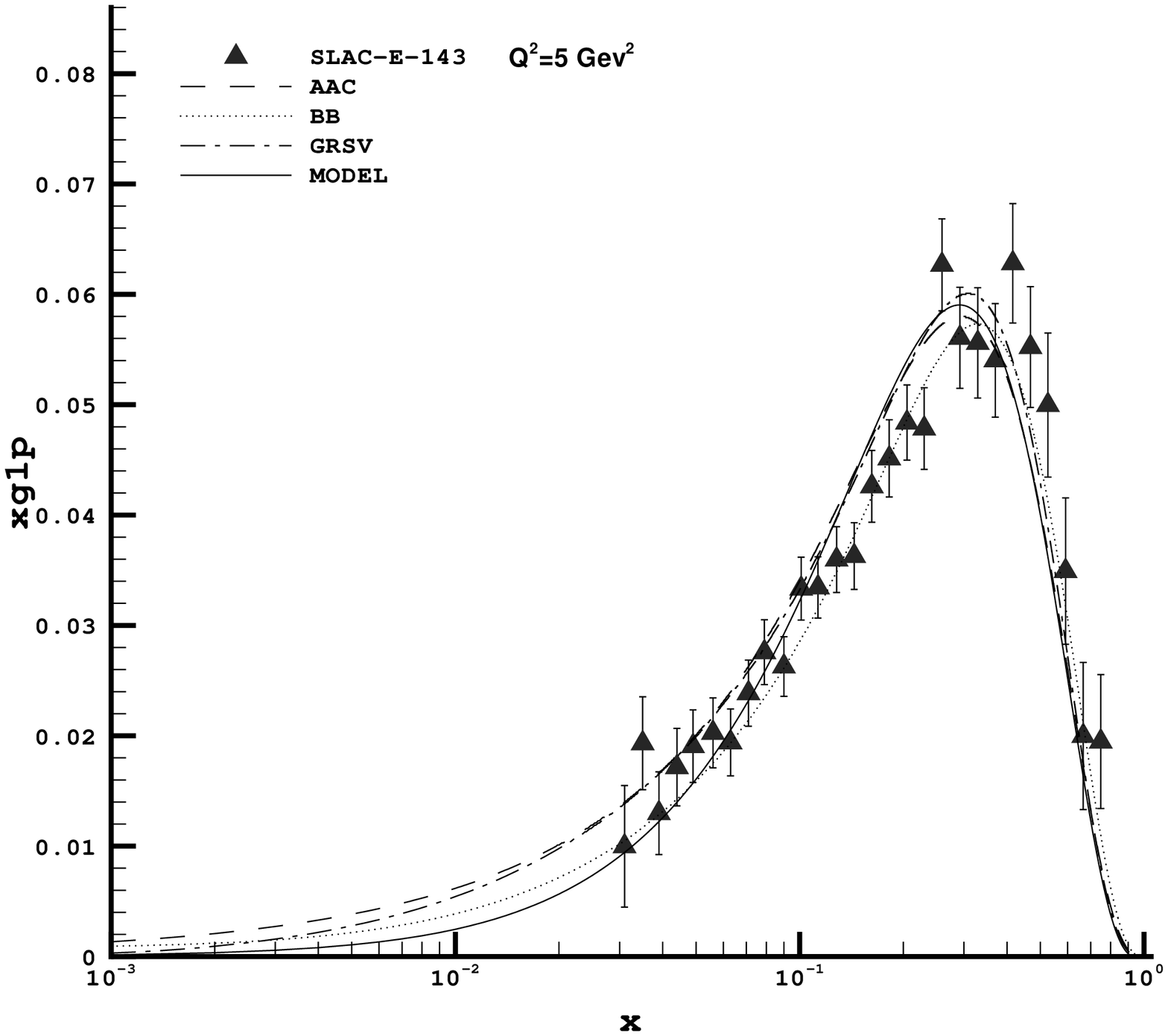,width=9cm} \\
\epsfig{figure=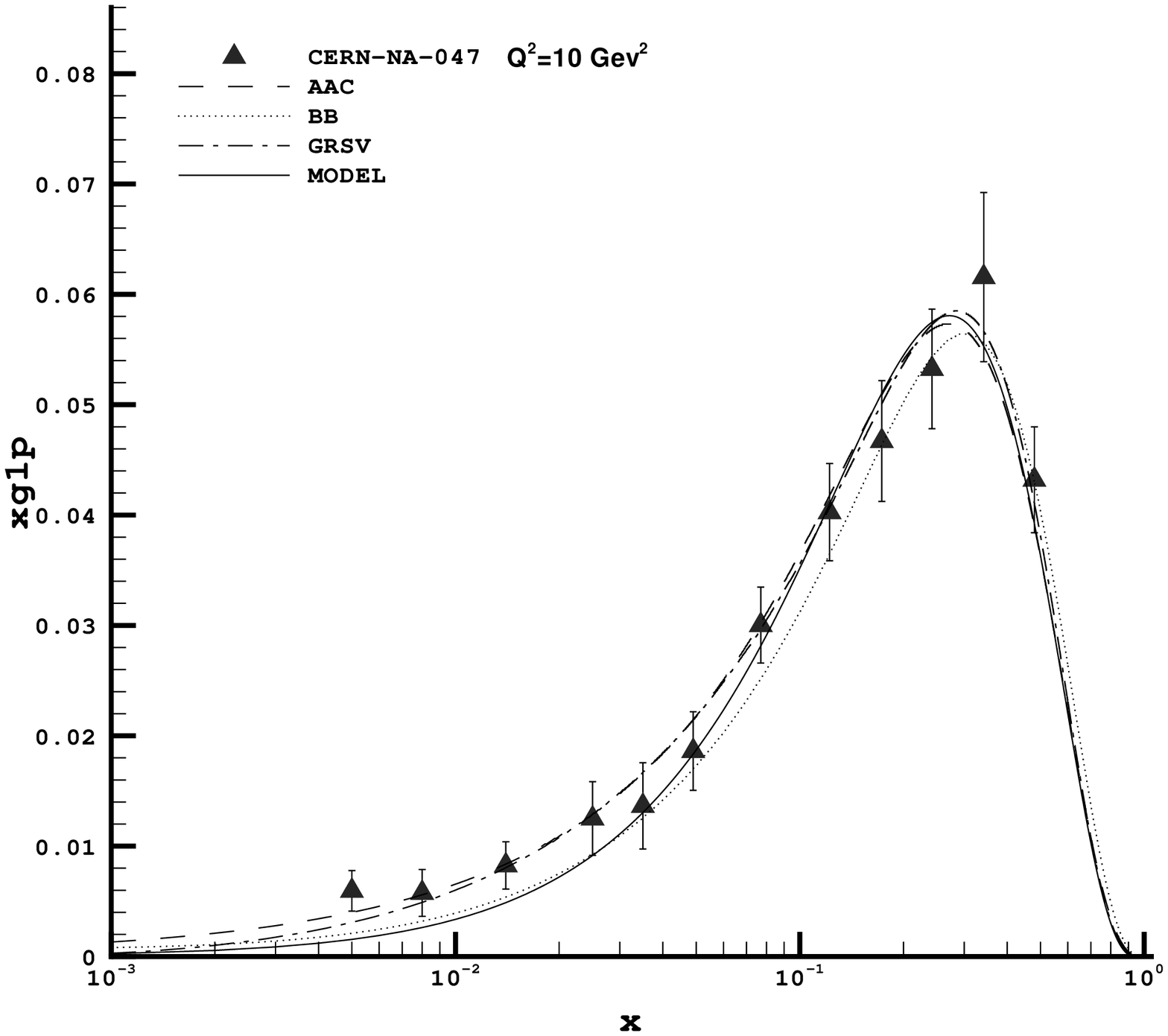,width=9cm}
\end{tabular}}
\caption{\footnotesize Polarized proton structure function,
$xg_{1}^{p}$, as a function of $x$ at various $Q^2$ values. Data
points are from [18, 19, 24,25] } \label{figure 5}
\end{figure}
It is evident that the model calculation is in good agreement with
the experimental data. In Fig. 6 we compare the calculated results
of $x\delta u_v (x)$ and $x\delta d_v (x)$ with the experimental
values from HERMES at $Q^2 =2.5$ $GeV^2.$ Results from other analysis
are also shown. \\
\begin{figure}
\centerline{\begin{tabular}{cc}
\epsfig{figure=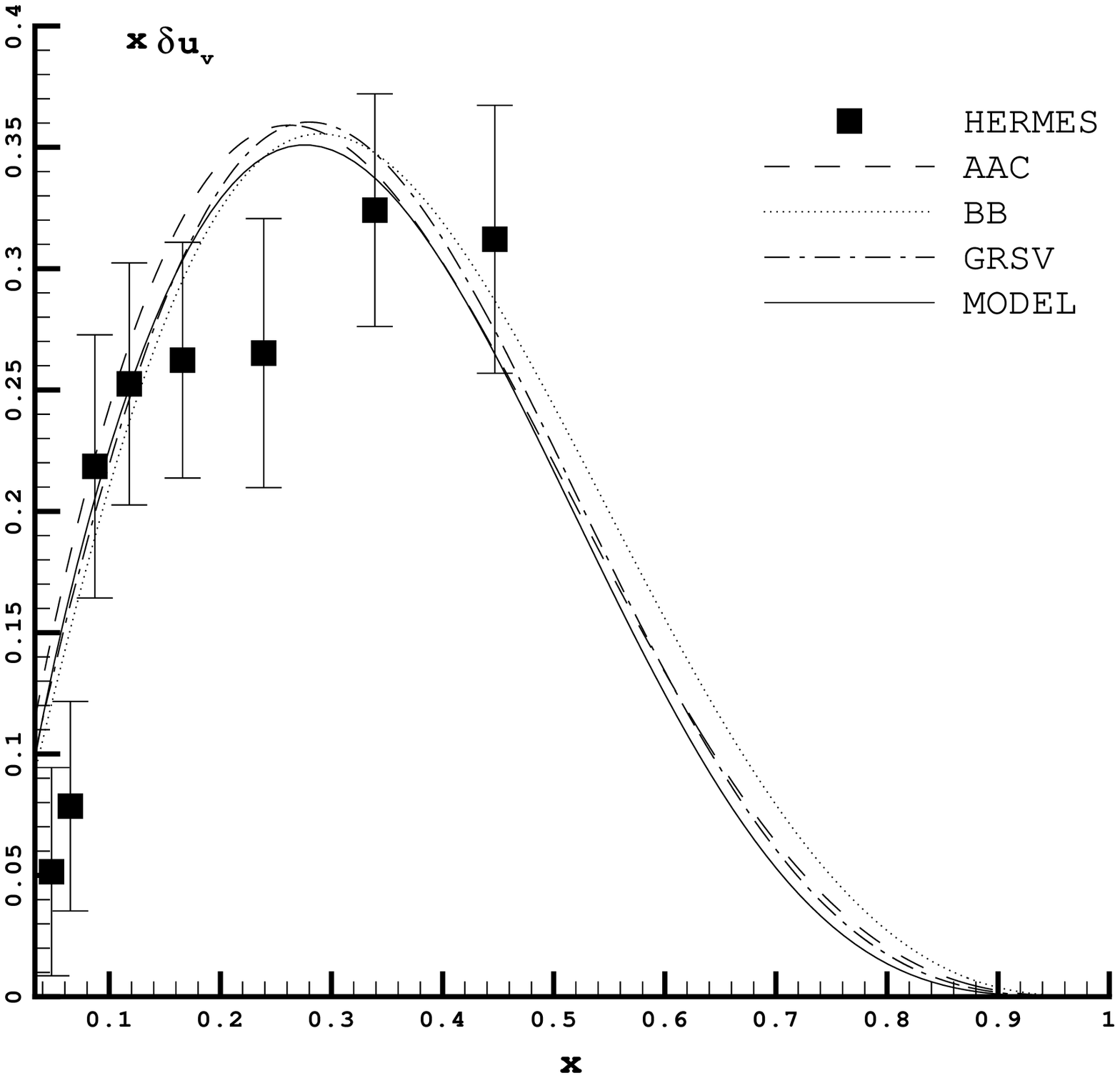,width=8cm}
\epsfig{figure=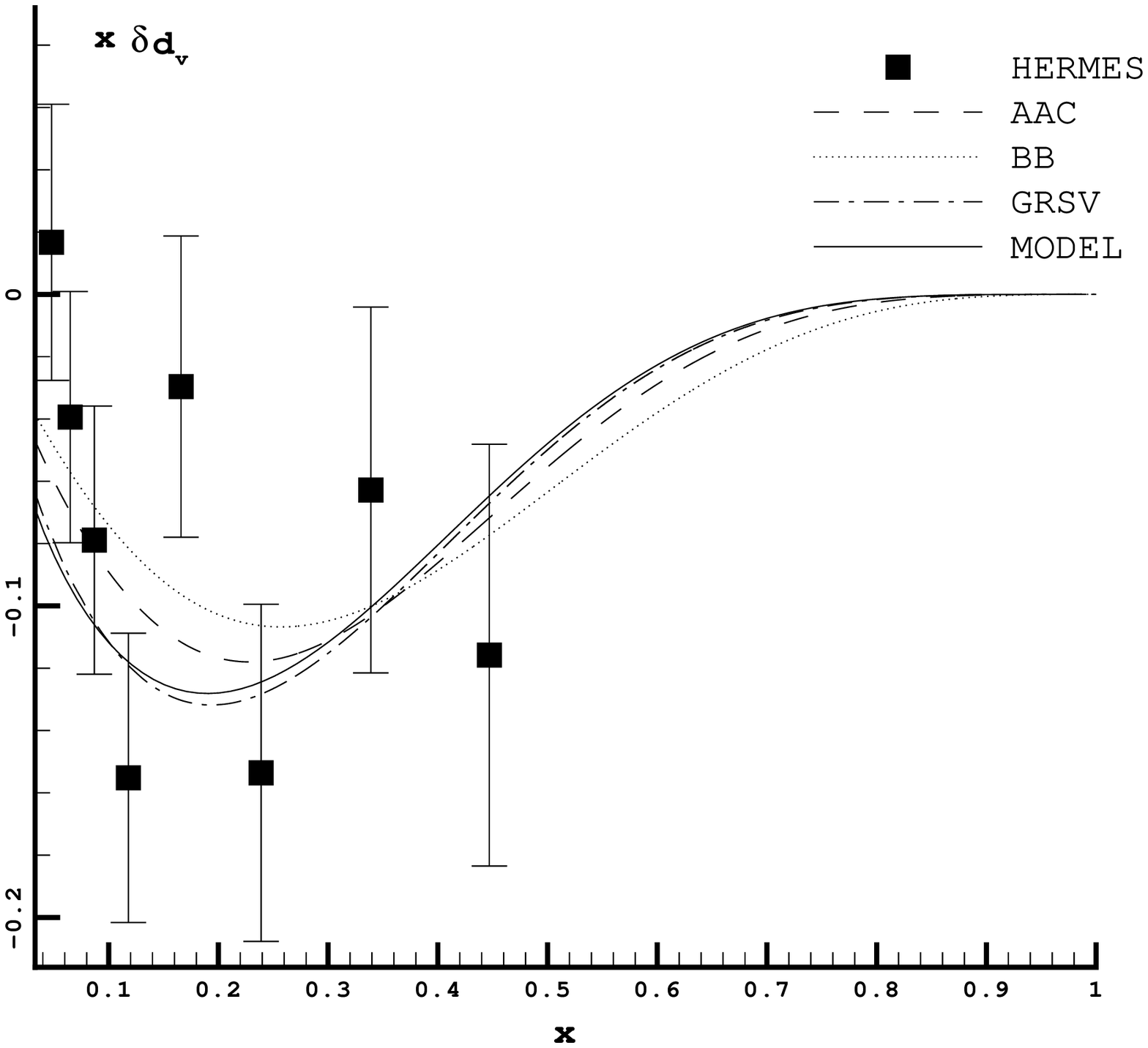,width=8cm}
\end{tabular}}
\caption{\footnotesize $x\delta u_v (x)$ and $x\delta d_v (x)$ at
$Q^2=2.5 $ $GeV^2$. The curves are the model results and the data
points are from Ref. [18, 19] } \label{figure 6.}
\end{figure}
As a further comparison, in Figs. 7 and 8 we present results for
$xg_{1}^{n}$ and $xg_{1}^{d}$. These results are also in agreement
with the experimental data \cite{24} and the analysis of
References \cite{16}, \cite{26}, and \cite{27}.
\begin{figure}
\centerline{\begin{tabular}{cccccccc}
\epsfig{figure=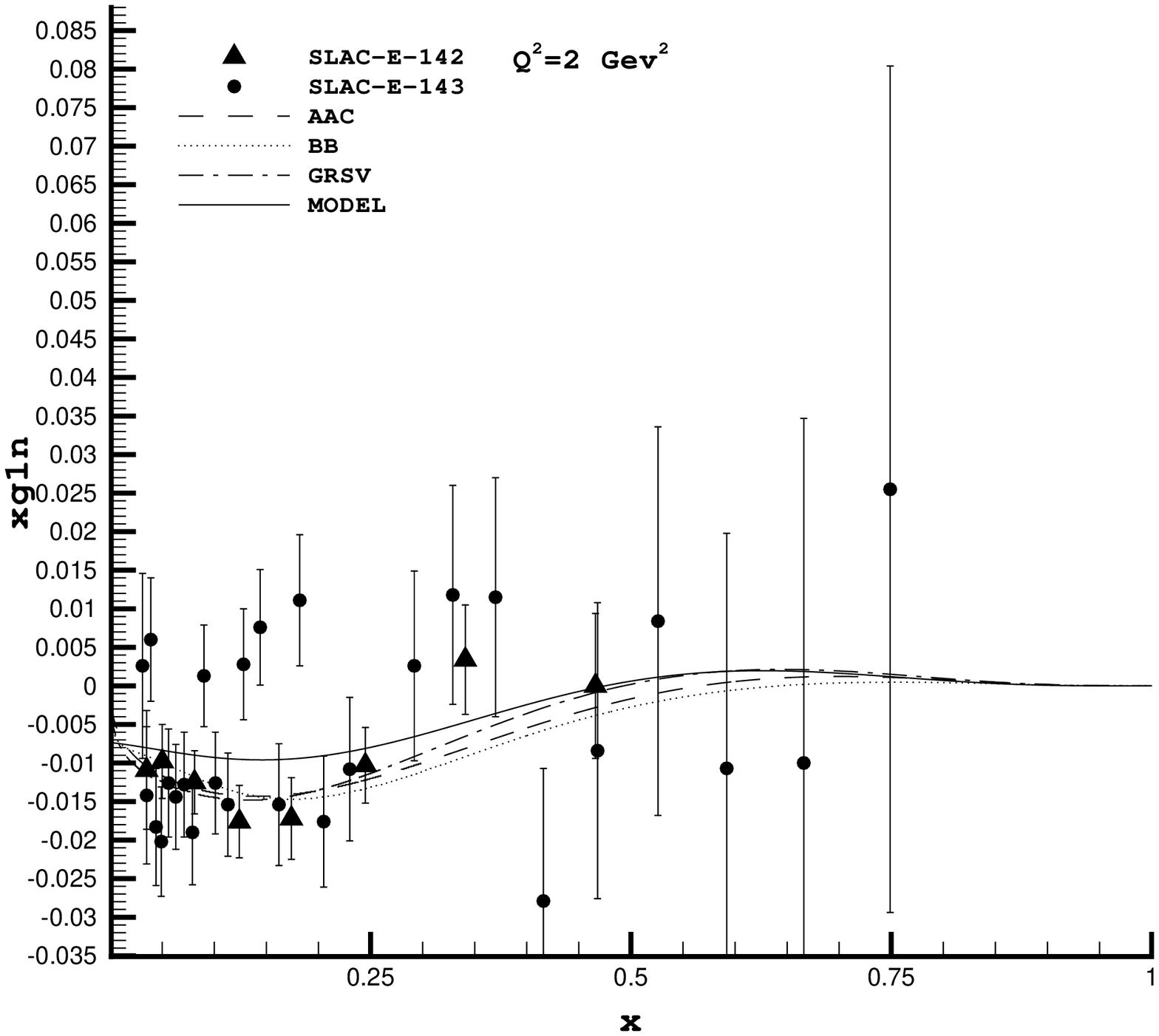,width=6cm}
\epsfig{figure=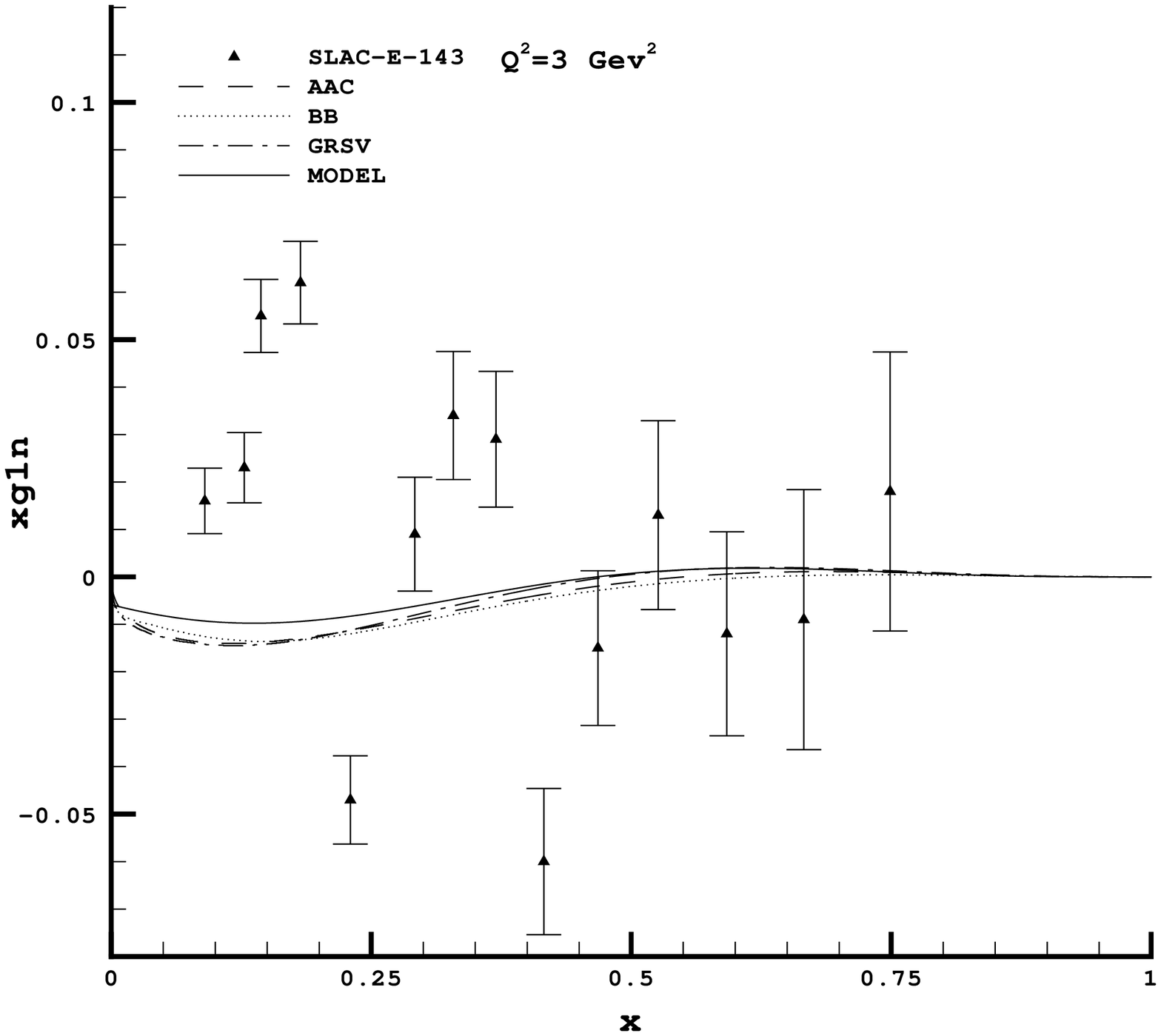,width=6cm}
\epsfig{figure=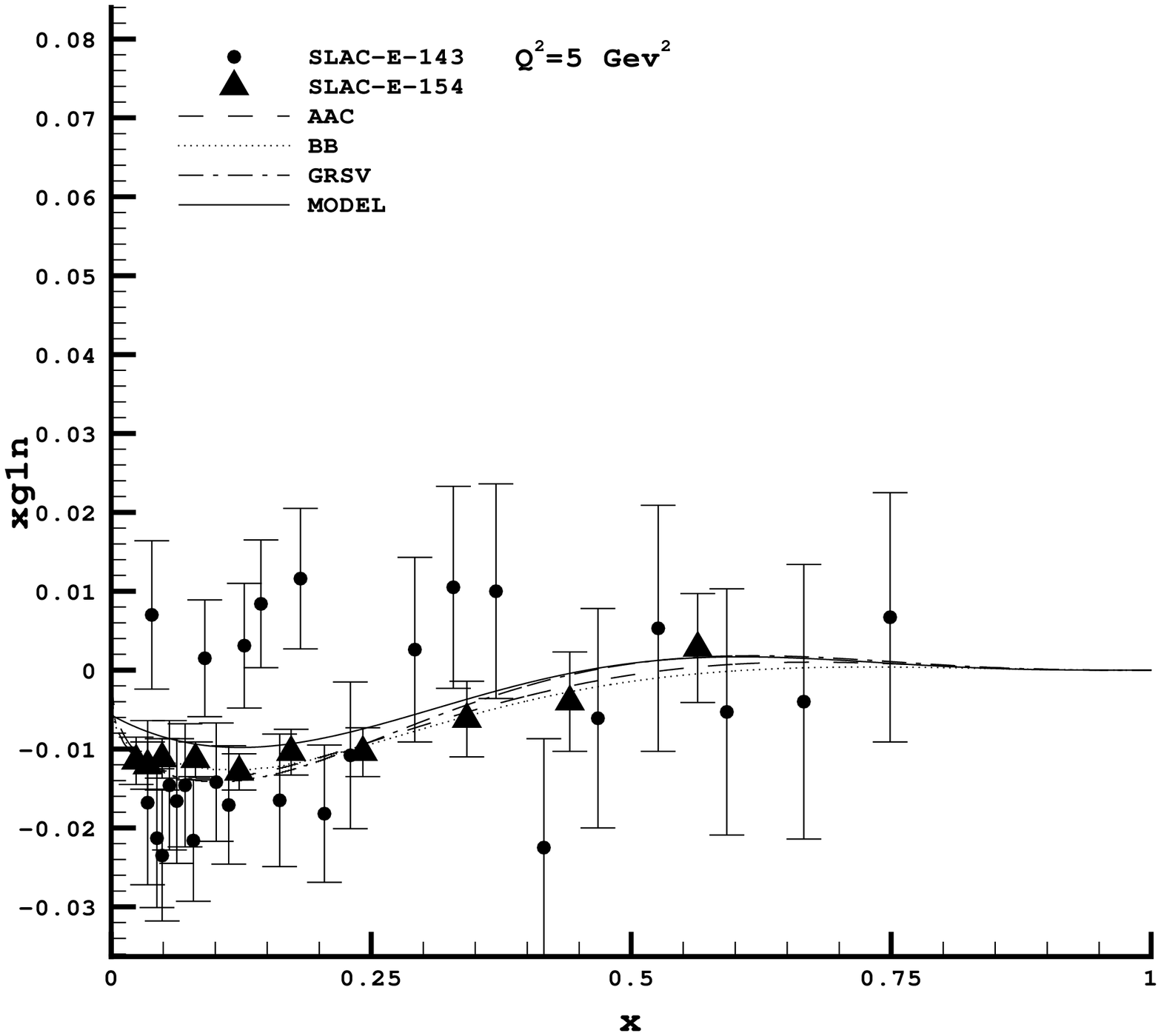,width=6cm}
\end{tabular}}
\caption{\footnotesize Polarized neutron structure function,
$xg_{1}^{n}$, as a function of $x$ at $Q^2=2, 3$, and $5$ $GeV^2$.
Data points are from \cite{24}. } \label{figure 7.}
\end{figure}
\\
\begin{figure}
\centerline{\begin{tabular}{cccccccc}
\epsfig{figure=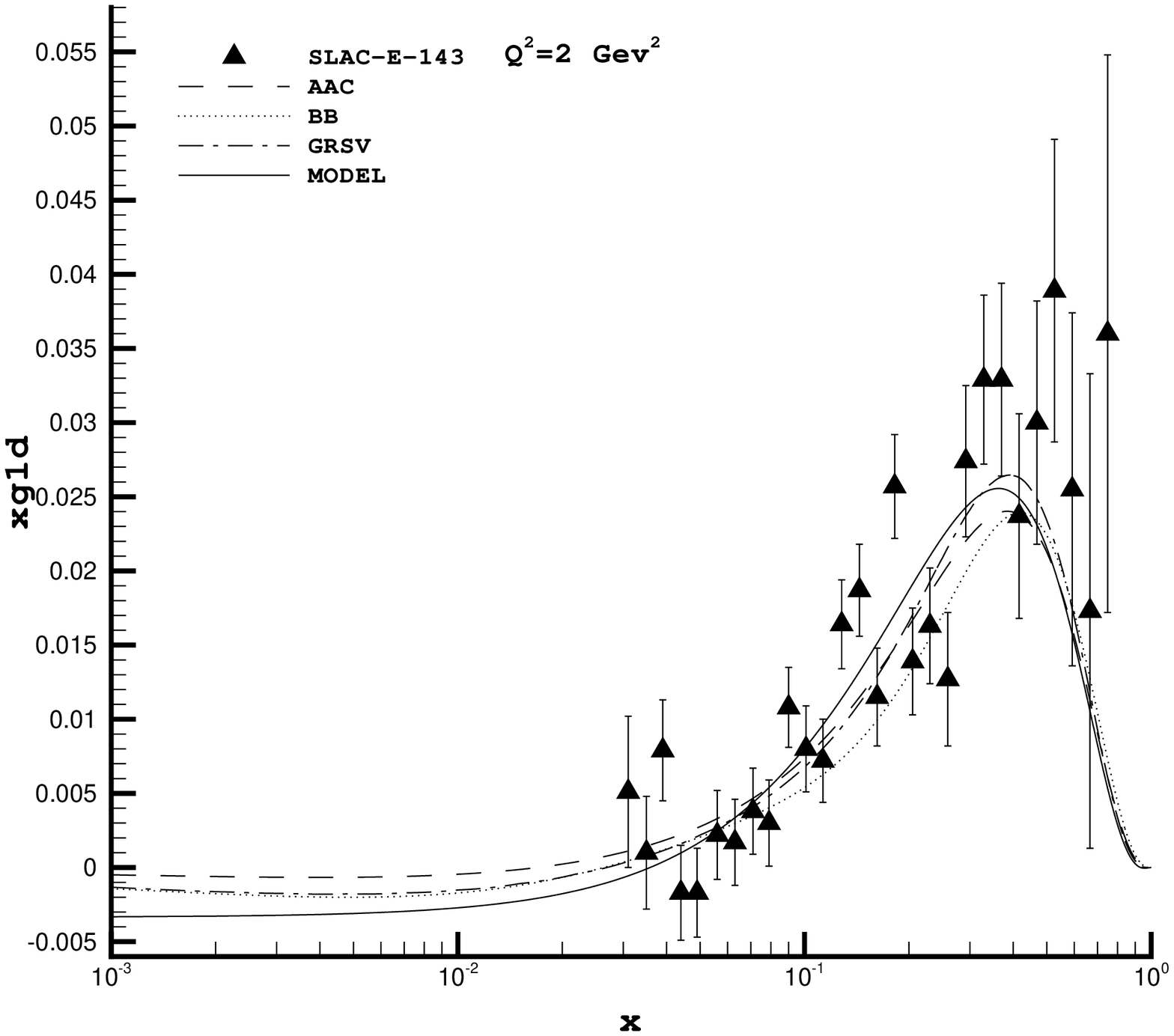,width=6cm}
\epsfig{figure=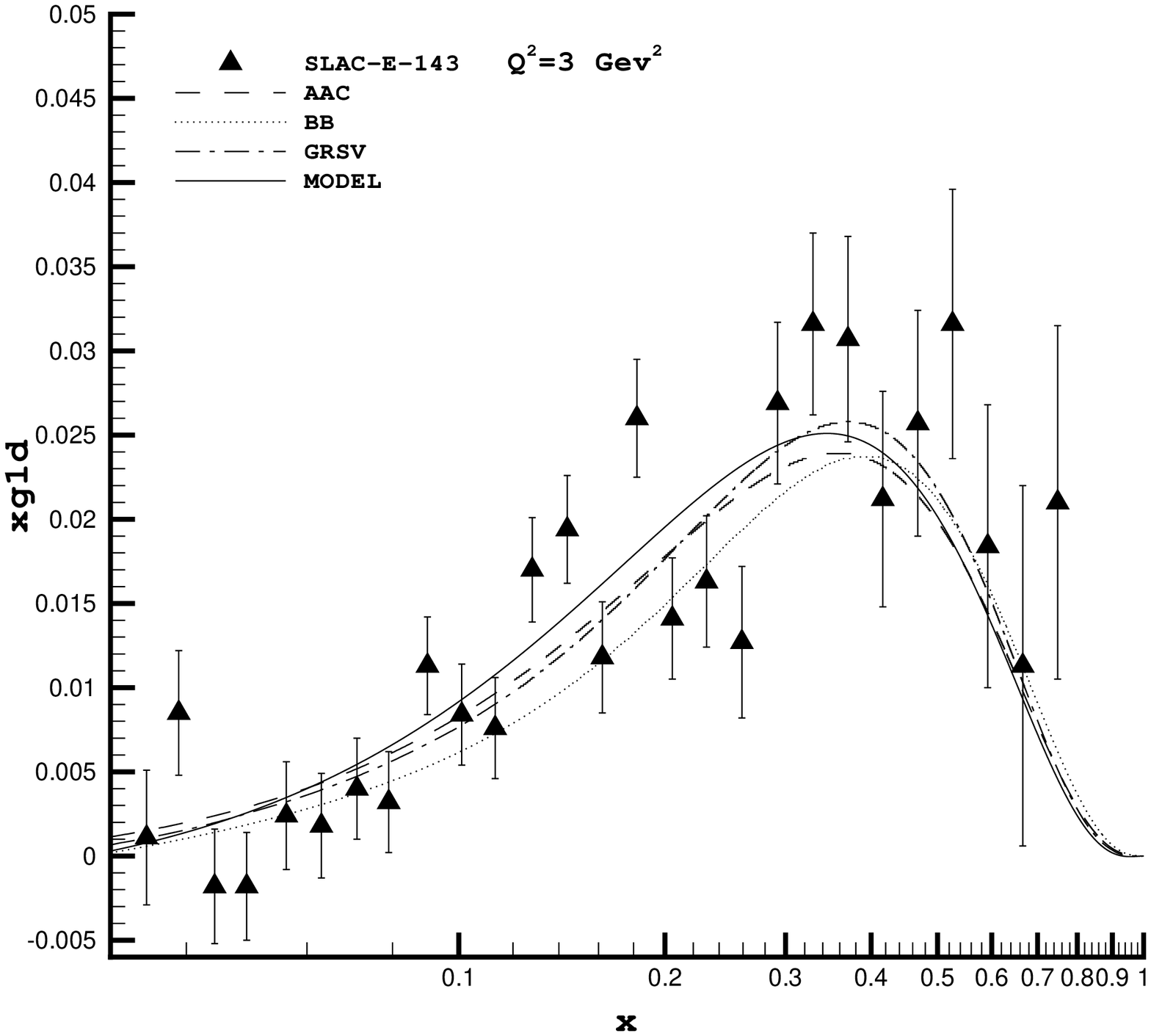,width=6cm}
\epsfig{figure=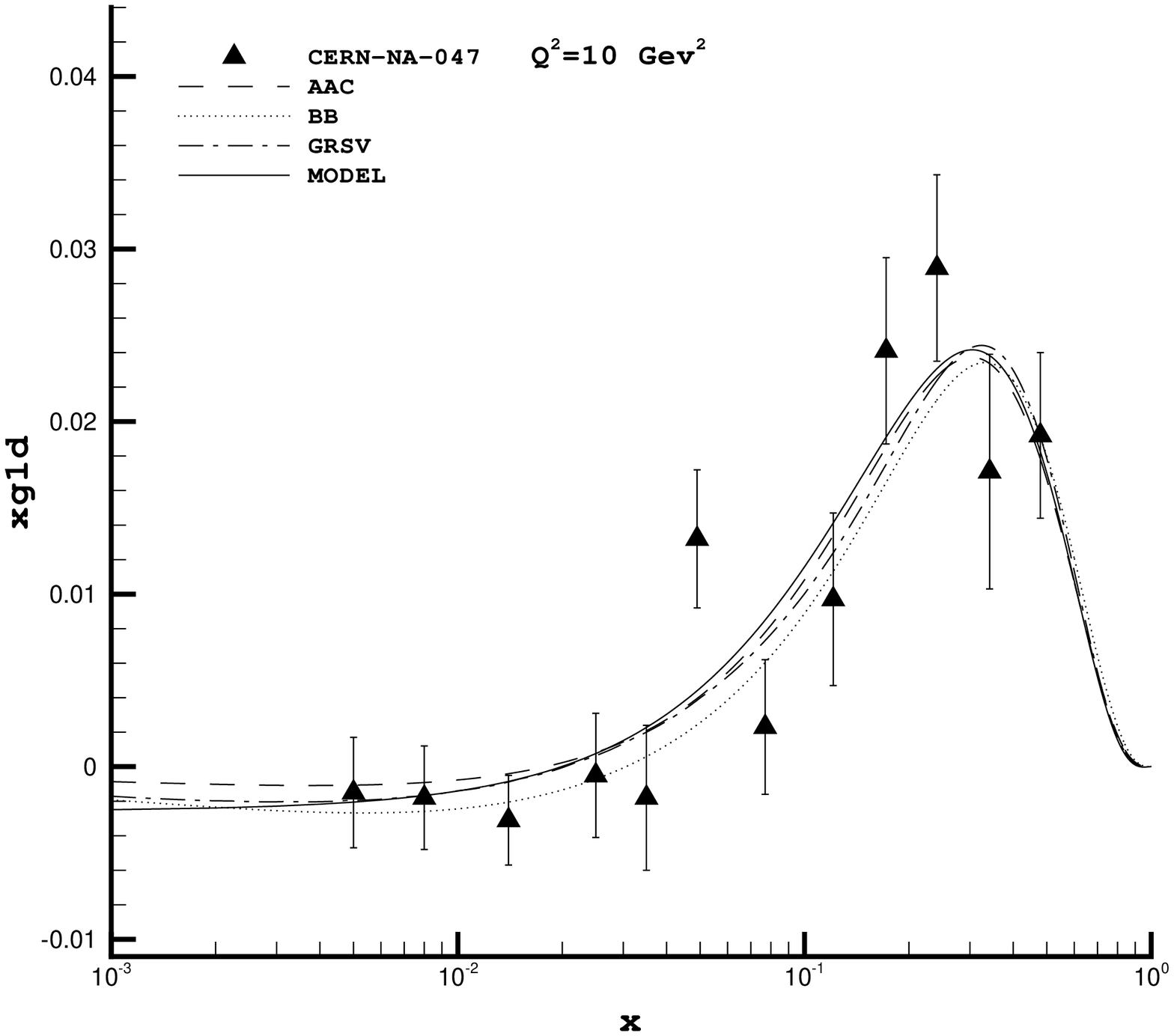,width=6cm}
\end{tabular}}
\caption{\footnotesize Polarized deuteron structure function,
$xg_{1}^{d}$, as a function of $x$ at $Q^2=2, 3$, and $10$
$GeV^2$. Data points are from \cite{24}. } \label{figure 8.}
\end{figure}
\section{First Moments and the Spin of Proton}
The first moment of polarized proton structure function, defined
by
\begin{equation}
\Gamma_{1}^{p}=\int_{0}^{1} g_{1}^{p}(x,Q^2)dx
\end{equation}
can be related to the combinations of the quark spin components
via
\begin{equation}
\Gamma_{1}^{p}=\frac{1}{2}\sum_{q} e^{2}_{q} \Delta
q(Q^2)=\frac{1}{2}\sum_{q} e^{2}_{q}< p,s\mid
\overline{q}\gamma_{\mu} \gamma_{5} q\mid p,s > s^{\mu}.
\end{equation}
Our results for $\Gamma^{p}_{1}$ are listed in Table II. The
moments of the polarized quark contributions to the spin of proton
at, say, $Q^2=3$ $GeV^2$ are
\begin{equation}
\Delta u_{valence}=0.820, \hspace {2cm} \Delta d_{valence}=-0.422,
\hspace{1.5 cm} \Delta \overline{q}_{sea}\sim 0
\end{equation}
For other values of $Q^2$ similar results are also obtained: for
$Q^2 =[2,10]$, we have $\Delta u_{v}=[0.816,0.827]$ and $\Delta
d_{v}=[-0.420,-0.426]]$.  These results for the first moment of
$g_{1}^{p,n,d}$, the quantity
$\Gamma^{p,n,d}_{1}$, yield the values that are presented in Table II. \\
{\it{Table II. Numerical values for $\Gamma^{N}_{1}$ at several $Q^2$ values}} \\
\begin{tabular}{ccccccccc}
\hline\hline $\Gamma_{1}^{N}$ & $Q^2=2 GeV^2$ & $Q^2=2.5 GeV^2$ &
$Q^2=3 GeV^2 $ &$ Q^2=5 GeV^2$ & $Q^2=10 GeV^2 $  \\
\hline
$p$ & 0.1132 & 0.1153 & 0.1168 & 0.1200 & 0.1245  \\
$n$ & -0.0630 & -0.0459 & -0.0554 & -0.0548 & -0.0546 & \\
$d$ &0.0305 & 0.0386 & 0.0341 & 0.0357 & 0.0377&
\\
\hline\hline
\end{tabular}
\\ \\
The available experimental values for $\Gamma^{p,n}_{1}$ are
obtained in a range of $Q^2$, rather than at a fixed $Q^2$ and a
fit to all data at $Q^2=5$ yields $\Gamma^{p}_{1}=0.118 \pm 0.004
\pm 0.007$ and $\Gamma^{n}_{1}=-0.048 \pm 0.005 \pm 0.005$
\cite{28}. More recent data from HERMES \cite{31} suggest that
$\Gamma^{p}_{1}=0.1211 \pm 0.005 \pm 0.008$ and
$\Gamma^{d}_{1}=0.0436 \pm 0.0012 \pm 0.0018$ at $Q^{2}=5$
$GeV{2}$. The values given in Table II matches these experimental
results.
\newline Additional information on the quark polarization
is also available from the low-energy nucleon axial coupling
constants $g_{A}^{3}$ and $g_{A}^{8}$:
\begin{eqnarray}
\begin{array}{c}
g_{A}^{3} \equiv <p,s \mid \overline{u}\gamma_{\mu} \gamma_{5}u -
\overline{d}\gamma_{\mu} \gamma_{5}d \mid p,s>s^{\mu} =\Delta
u(Q^2)-\Delta d(Q^2), \\
g_{A}^{8} \equiv <p,s \mid \overline{u}\gamma_{\mu} \gamma_{5}u +
\overline{d}\gamma_{\mu} \gamma_{5}d -2 \overline{s}\gamma_{\mu}
\gamma_{5}s\mid p,s>s^{\mu}=\Delta u(Q^2)+\Delta d(Q^2)-2 \Delta
s(Q^2).
\end{array}
\end{eqnarray}
Since there is no anomalous dimension associated with the
axial-vector currents, $A_{\mu}^{3}$ and $A_{\mu}^{8}$, the
non-singlet couplings, $g_{A}^{3}$ and $g_{A}^{8}$ do not evolve
with $Q^2$ and hence can be determined from low-energy neutron and
hyperon $\beta$-decays. The experimental values are
$g^{3}_{A}=1.2573 \pm 0.0028$ and $g^{8}_{A}=0.579 \pm 0.025$. The
constraining values of $g^{3,8}_{A}$ are the ones used in most
analysis performed in order to fix sea quark contributions, in
particular that of $\Delta S$. We have not considered this
restriction {\it{a priori}}; instead we want to see if the results
of the model can reproduce these numbers. From the stated values
for $\Delta u_{v}$ and $\Delta d_{v}$ we obtain $g^{3}_{A}=
1.240-1.253$ which accommodates the experimental value with an
accuracy of $2\%$. Our agreement with $g^{8}_{A}$ is not as good
as $g^{3}_{A}$. However, this does not invalidate the results of
the model, for there are serious objections \cite{29} \cite{30} (
mainly due to $m_{u,d}\ll m_{s}$) to
\begin{equation}
g^{8}_{A}=\Delta u +\Delta \overline{u}+ \Delta d +\Delta
\overline{d}-2(\Delta s +\Delta \overline{s})=3F-D=0.579 \pm
0.025,
\end{equation}
in contrast to the unquestioned isospin $SU(2)$ symmetry
($m_{u}\simeq m_{d}$) that gives rise to the value of $g^{3}_{A}$.
Experimental data from HERMES also puts the value of $g^{8}_{A}$
at $0.274 \pm 0.026 \pm 0.011$ in the range of $0.02< x < 0.6$
\cite{31} which is substantially less than the value inferred from
hyperon decay. We have obtained $g^{8}_{A}= 0.39$. Findings of
\cite{31} also suggests that $\Delta s + \delta {\overline{s}}$ is
consistent with zero. The HERMES data \cite{31} in the measured
region of $x> 0.02$ gives $\Delta s + \delta {\overline{s}}=
0.006\pm 0.029 \pm 0.007$. If the results reported in second
reference of \cite{31} is confirmed, it would also rule out any
significant non-perturbative effects that predict strange quark
contribution to the spin of proton from pion-nucleon sigma term
$\sigma_{\pi N}$ which in turn alters the value of $g^{8}_{A}$.
Such a contribution can be calculated in the framework of chiral
quark model. In fact, in this framework it is shown that \cite{32}
while light sea quark polarization is consistent with zero, the
$\Delta S=-0.051$. If it is added to our result,it becomes in line
with $g^{8}$ value obtained from hyperon decay. Nevertheless, The
crucial point in our model is that there is no room for the
perturbative sea quark polarization, because the sea of the valon
is generated entirely from gluon splitting and for the massless
quark, the helicity is conserved; and yet the model nicely
accommodates all of the experimental data with acceptable
accuracy. Our finding that $\Delta q_{sea}$ is consistent with
zero is in agreement with the HERMES \cite{18} \cite{19} and SMC
collaboration \cite{33} data.
This can be seen yet in a different way as follows:\\
The moments of sea quarks in a proton can be written as
\begin{equation}
\Delta \overline{q}(n, Q^{2})=\frac{1}{2f}[\Delta \Sigma(n,
Q^{2})-\Delta u_{v}(n, Q^{2})-\Delta d_{v}(n, Q^{2})]
\end{equation}
where
\begin{eqnarray}
\begin{array}{c}
\Delta u_{v}(n,Q^{2})=2 \Delta M^{NS}(n, Q^{2})\otimes \Delta
M_{\frac{U}{p}}(n) \\
\Delta d_{v}(n,Q^{2})=\Delta M^{NS}(n, Q^{2})\otimes \Delta
M_{\frac{D}{p}}(n)\\
\Delta \Sigma(n, Q^{2})=\Delta M^{S}(n, Q^{2})\otimes [2\Delta
M_{\frac{U}{p}}(n)+ \Delta M_{\frac{D}{p}}(n)]
\end{array}
\end{eqnarray}
where $\Delta M_{\frac{U,D}{p}}(n)$ are moments of $\delta
G_{U,D}$. For $n=1$ they are given by
\begin{equation}
\int^{1}_{0} dy \delta G_{\frac{U}{P}}(y)=0.403,
\hspace{2cm}\int^{1}_{0} dy \delta G_{\frac{D}{P}}(y)=-0.409
\end{equation}
Therefore, EQ.(25)becomes
\begin{equation}
\Delta\overline{q}(n, Q^2)=(\Delta M^{S}(n,Q^2) - \Delta
M^{NS}(n,Q^2)) (2\Delta M_{\frac{U}{p}}(n,Q^2)+ \Delta
M_{\frac{D}{p}}(n,Q^2))
\end{equation}
and we see that for $n=1$ the range of variation for $\Delta
\overline{q}(1, Q^{2})$ is $0 - 0.016$, since $\Delta M^{S}(1,
Q^{2})=1$ and $\Delta M^{NS}(n=1)$ varies between 1 and 0.80 for
$Q^{2}=[0.283 , 10^{6}]$ $GeV^{2}$; that is, the contribution of
the sea quark to the spin of proton is consistent with zero.
Reference [15] introduces two different polarized valon
distribution in an attempt to avoid dealing with $\Delta
\overline{q}(1, Q^{2})\approx0$. Such an scheme is quite counter
intuitive, because it means that singlet and non-singlet quark
distribution inside a valon alters the valon distribution in a
hadron. A notion
that is hard to understand within the context of the valon model. \\
Of course a $2\%$ deviation of our results from the value of
$g^{3}_{A}$ can be attributed to the presence of sea polarization
which then ought to be generated non-perturbatively . This point
is not considered here nor have we attempted to fit the data in
order to extract possible sea quark contribution, in particular
that of strange sea. Another possible source for this deviation
can be a poor determination of $\delta G_{\frac{U, D}{p}}$. We
also note that a plausible alternative to the full $SU(3)_{f}$
symmetry is a "valence" scenario where $SU(3)_{f}$ symmetry is
maximally broken which is based on the assumption that the
flavor-changing hyperon $\beta$-decay data fix only the total
helicities of valence quark at some appropriately chosen input
scale $Q^2 =Q^2_{0}$ \cite{17}. \newline Our prediction for
$\Delta \Sigma$, the total quark contribution to the spin of
proton, lies in the range of $0.410-0.420$ for $Q^2=[2,10]$. The
variation of $\Delta \Sigma$ is due to (marginal) $Q^2$ dependence
of $\Delta q_{v}$ in the Next-to-Leading Order; because $\Delta
P^{(1)}_{NS}\neq 0$. This result is also compatible with the
experimental data of Refs. [17, 18], where for the measure range,
$0.023< x <0.6$, they obtained $\Delta \Sigma =0.347 \pm 0.024 \pm
0.066$. Further experimental support for our findings comes from
recently published data from COMPASS collaboration \cite{34} at
$Q^{2}=10$ $GeV^{2}$ and $0.006<x<0.7$. It shows that $\Delta
u_{v} + \Delta d_{v}=0.40 \pm 0.07\pm 0.05$ , $\Delta
{\overline{u}} + \Delta {\overline{d}}=0.0 \pm 0.04 \pm 0.03$, and
$\Delta{s} + \Delta {\overline{s}}=-0.08 \pm 0.01 \pm0.03$. This
result is consistent with $\Delta {\overline{u}}= \Delta
{\overline{d}}$. The conclusion is that: if we accept the validity
of both HERMES \cite{31} and COMPASS \cite{34} data, then the role
of sea quark polarization in calculation of polarized structure
functions, $g_{1}^{p, n, d}$ is marginal. We have arrived at the
same conclusion directly by considering only the QCD processes up
to the Next-to-Leading order in the valon representation.

\subsection{Role of Gluon, Orbital angular momentum, and the spin of proton}
The phenomenological model that is described here is able to
account for the experimental data with a good accuracy. However,
it still remains to accommodate the spin of proton. The spin of
valon in the absence of gluon is accounted for by the total spin
contribution of quarks, as we saw in section 2. The large gluon
polarization in valon, however requires a sizable negative orbital
angular momentum to compensate for the gluon contribution beyond
$Q_{0}^{2}$. The implication for proton is that $\Delta g$ rises
as $Q^2$ increases; being around 0.4 at $Q^2=2$ $GeV^{2}$ and
reaches to 0.7 at $Q^2=14$ $GeV^{2}$. For the same range of $Q^2$,
then $L_{z}$ varies from $-0.1$ to $-0.4$. These results are
plotted in Figure 9. The role of Orbital angular momentum, $L_{z}$
in a valon is to cancel out the gluon polarization completely, but
this cancelation in proton is partial. Therefore, it is reasonable
to speculate that about $60\%$ of the net spin of proton comes
from gluon. This is comparable with $\sim 50 \% $ momentum
contribution of gluon to the total momentum of proton. We do share
the opinion expressed in \cite{31} based on the experimental data
from HERMES that the quark helicities contribute a substantial
fraction to the nucleon helicity, but there is still need for a
major contribution from gluon/orbital angular momentum. We have
concluded that the orbital angular momentum is needed and arises
from the structure of valon in order to compensate for the growing
gluon helicity and produce a spin-$\frac{1}{2}$ valon. It also can
be argued that the valons of nucleon themselves might have a
relative orbital angular momentum other than zero. Such a
situation would amount to the assumption that the nucleon is not
in the $S$- state of orbital angular momentum, the case that we
have not considered in this analysis. In the analysis of section 2
we stated that the magnitude of gluon helicity, $\Delta g$,
depends on the initial scale, $Q^{2}_{0}$, chosen; and hence, so
does the values of $L_{z}$. We believe that the mathematical
boundary condition of the model provides a reasonable guideline
for the choice of $Q^{2}_{0}$.
\begin{figure}
\centerline{\begin{tabular}{cccccccc}
\epsfig{figure=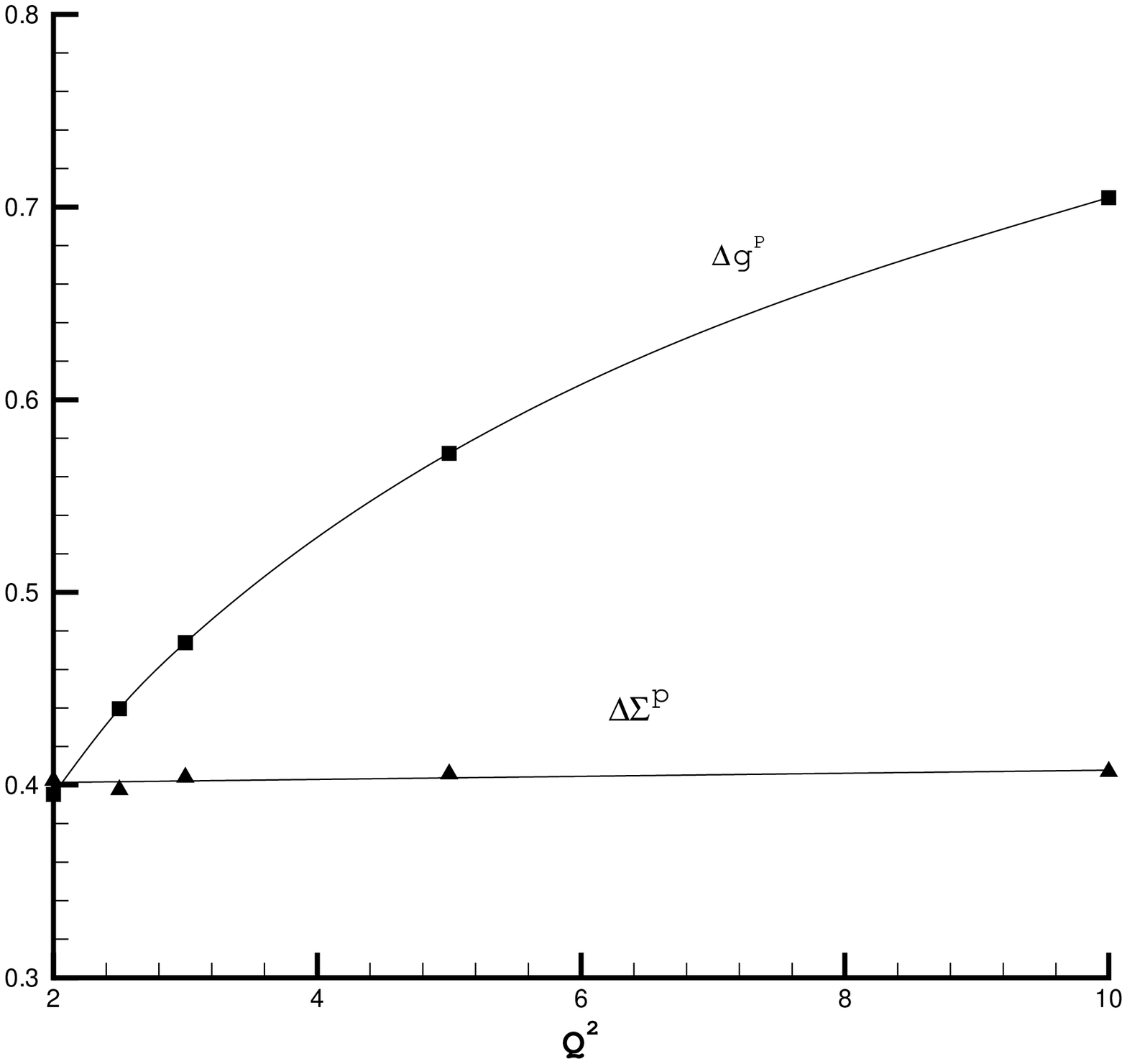,width=7cm} \epsfig{figure=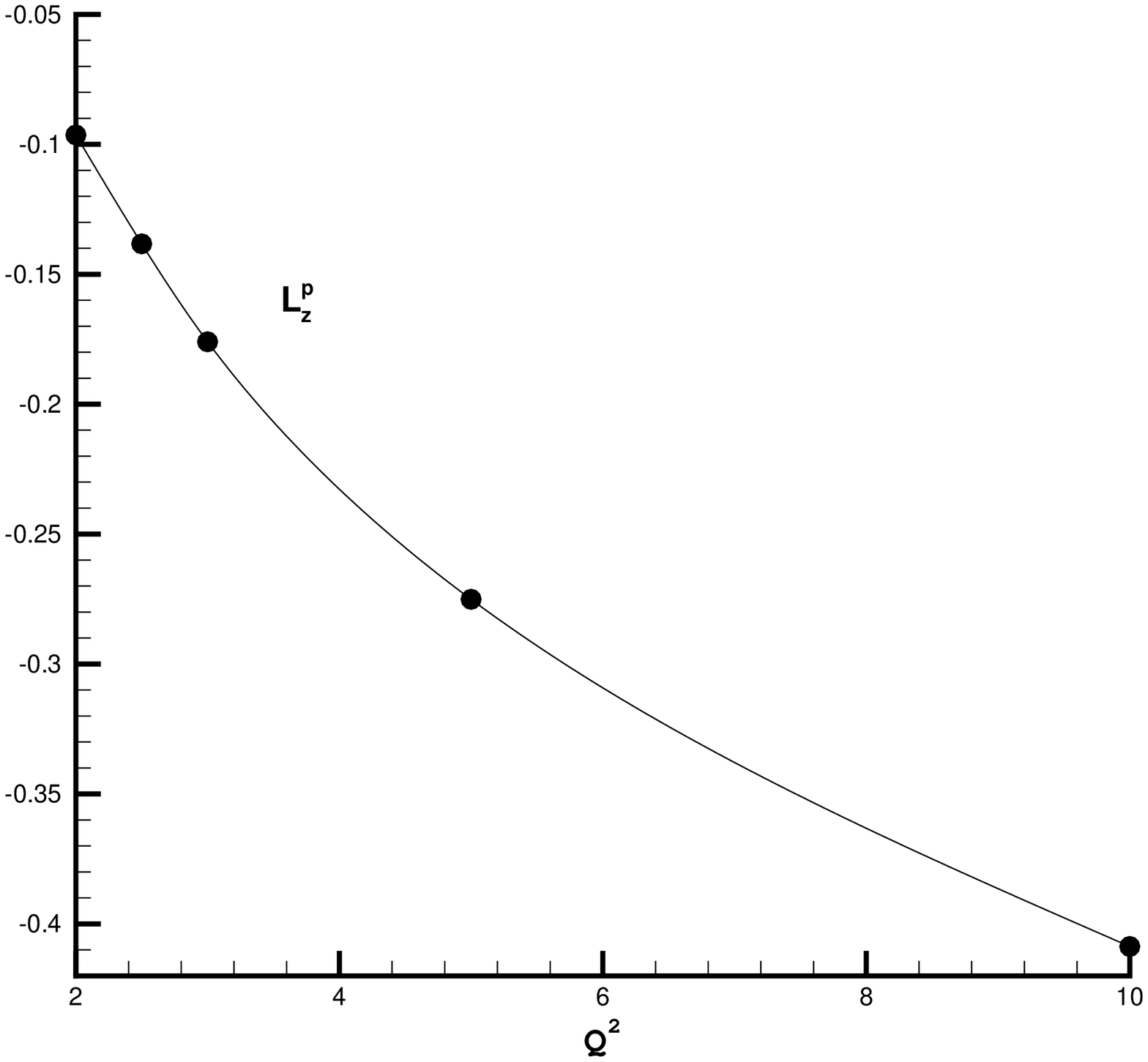,width=7cm}
\end{tabular}}
\caption{\footnotesize Gluon, $\Delta g$, and orbital angular
momentum , $L_{z}$, components in proton as a function of
$Q^{2}$.} \label{figure 9.}
\end{figure}

\section{Remarks and conclusions}
We have calculated the polarized structure of a valon form QCD
processes in the next-to-leading order framework. While the
valence quark completely accounts for the spin of a valon, the
presence of large gluon polarization in the valon  makes it more
complicated, requiring a sizable and negative orbital angular
momentum. Our finding indicates that the sea parton polarization
in the valon remains very small, and hence its contribution to the
spin structure of proton is consistent with zero. This finding is
in agreement with the experimental results \cite{18}, \cite{19}.
The picture presented here is capable of reproducing all available
data on $g^{p,n,d}_{1}(Q^2)$ with good accuracy. We have further
calculated the orbital angular momentum contribution, and its
evolution, to the spin content of proton and valon. It appears
that the size of gluon contribution to the spin content of proton
is around $60 \% $, somewhat similar to the momentum contribution
of gluon to the momentum of proton in the unpolarized case. This
value is also sensitive to the initial scale, $Q_{0}^{2}$. The
model presented here does not have any free parameter, it is free
of data fitting and solely relies on QCD processes, except the use
of phenomenological concept of the valon model. Finally, we stress
that there is the issue of initial input densities at scale,
$Q^{2}_{0}$. While the most theoretical analysis and global fits
begin with $Q^{2}_{0}\geq 1$ $GeV^{2}$, we have used the
mathematical boundary conditions of the model and have shown that
the results are compatible with the experimental data.
\section{Acknowledgment} We are grateful to Professor
Guido Altarelli for an interesting discussion and his valuable
comments about the manuscript. We are also deeply appreciative of
Professor Mauro Anselmino for his critical reading of the
manuscript and his constructive comments.

\end{document}